\documentclass[12pt]{article}
\usepackage{enumerate}
\usepackage{amsfonts}
\usepackage{amsmath}
\usepackage{amssymb}
\usepackage{amsthm}

\def\H{\mathcal{H}}
\def\K{\mathcal{K}}
\def\P{\mathcal{P}}
\def\S{\mathfrak{S}}

\def\T{\mathfrak{T}}
\def\B{\mathfrak{B}}

\newcounter{defin}  \newcounter{lemma}  \newcounter{theorem}
\newcounter{property} \newcounter{corol}  \newcounter{remark} \newcounter{example}

\newenvironment{lemma}{\par\refstepcounter{lemma}%\noindent
     \textbf{Lemma \thelemma.} }{\rm\par}
\newenvironment{theorem}{\par\refstepcounter{theorem}%\noindent
     \textbf{Theorem \thetheorem.}\ }{\rm\par}
\newenvironment{property}{\par\refstepcounter{property}%\noindent
     \textbf{Proposition \theproperty.}\ }{\rm\par}
\newenvironment{corollary}{\par\refstepcounter{corol}%\noindent
     \textbf{Corollary \thecorol.} }{\rm\par}
\newenvironment{definition}{\par\refstepcounter{defin}%\noindent
     \textbf{Definition \thedefin.}\ }{\rm\par}
\newenvironment{remark}{\par\refstepcounter{remark}%\noindent
     \textbf{Remark \theremark.}}{\rm\par}

\newcommand{\id}{\mathrm{Id}}
\newcommand{\Tr}{\mathrm{Tr}}
\newcommand{\ri}{\mathrm{ri}}

\begin{document}
\title{Reversibility of a quantum channel: general conditions and their applications to Bosonic linear channels}

\author{M.E.~Shirokov\footnote{email:msh@mi.ras.ru}\\
Steklov Mathematical Institute, Moscow, Russia}

\date{} \maketitle

\vspace{-25pt}

\begin{abstract}
The method of complementary channel for analysis of reversibility
(sufficiency) of a quantum channel with respect to families of input
states (pure states for the most part) are considered and applied to
Bosonic linear (quasi-free) channels, in particular, to Bosonic
Gaussian channels.

The obtained reversibility conditions for Bosonic linear channels
have  clear physical interpretation and their sufficiency is also
shown by explicit construction of  reversing channels. The method of
complementary channel gives possibility to prove necessity of these
conditions and to describe all reversed families of pure states in
the Schrodinger representation.

Some applications in quantum information theory are considered.

Conditions for existence of discrete classical-quantum subchannels and of completely depolarizing
subchannels of a Bosonic linear channel are obtained in the Appendix.
\end{abstract}

\vspace{-15pt}

\tableofcontents

\section{Introduction}

Reversibility (sufficiency) of a quantum channel
$\Phi:\mathfrak{S}(\mathcal{H}_A)\rightarrow\mathfrak{S}(\mathcal{H}_B)$
with respect to a family $\S$ of states in
$\mathfrak{S}(\mathcal{H}_A)$ means existence of a quantum channel
$\Psi:\mathfrak{S}(\mathcal{H}_B)\rightarrow\mathfrak{S}(\mathcal{H}_A)$
such that $\Psi(\Phi(\rho))=\rho$ for all $\rho\in\S$ \cite{J-rev,P-sqc}.

The notion of reversibility of a channel naturally arises in
analysis of different general questions of quantum information
theory, in particular, of conditions for preserving  entropic
characteristics of quantum states under action of a channel
\cite{H&Co,J&P,J-rev,O&Co,P-sqc}. For instance, it follows from
Petz's theorem (cf. \cite{J&P,P-sqc}) that the Holevo
quantity\footnote{The Holevo quantity (defined in Section II)
provides an upper bound for accessible classical information which
can be obtained by applying a quantum measurement
\cite{H-SCI,N&Ch}.} of an ensemble $\{\pi_i,\rho_i\}$ of quantum
states is preserved under action of a quantum channel $\Phi$, i.e.
$$
\chi(\{\pi_i,\Phi(\rho_i)\})=\chi(\{\pi_i,\rho_i\}),
$$
if and only if the channel $\Phi$ is reversible with respect to the
family $\{\rho_i\}$.

A general criterion for reversibility of a quantum channel expressed
in terms of von Neumann algebras theory is obtained in \cite{J&P}
(see also \cite{J-rev}). By using this criterion and the notion of
complementary channel  conditions for reversibility of a quantum
channel  with respect to complete families of states, in particular,
of pure states are obtained in \cite{Sh-RC}. These conditions can be
specified and reformulated for analysis of reversibility with
respect to noncomplete families. Moreover, their "necessary" parts
can be expressed in terms of weak complementary channel by using the
"face property" of a set of all channels reversible with respect to
a given family of states. These generalizations and their
corollaries, in particular, several criteria for reversibility of a
channel with respect to families of orthogonal mixed states are
considered in Section III.

In Section IV we apply these conditions to Bosonic linear
(quasi-free) channels. We show that a noisy Bosonic linear channel
is reversible neither with respect to any complete family of pure
states nor with respect to any complete family of orthogonal states
containing a finite rank state (but it may be reversible with
respect to  complete family of orthogonal infinite rank states).
Then we focus attention on analysis of reversibility of such
channels with respect to noncomplete orthogonal and nonorthogonal
families. The obtained conditions are reformulated for Bosonic
Gaussian channels playing a central role in quantum information
theory \cite{Caruso,E&W,H-SCI,Ivan}.

The obtained results imply conditions for
(non)\nobreakdash-\hspace{0pt}preserving  the Holevo quantity of
arbitrary (discrete or continuous) ensembles of states under action
of a quantum channel. They are considered in Section V.

\section{Preliminaries} Let $\H$ be a separable Hilbert space, $\B(\H)$ and
$\mathfrak{T}( \mathcal{H})$ -- the Banach spaces of all bounded
operators in $\mathcal{H}$ and of all trace-class operators in $\H$
correspondingly, $\S(\H)$ -- the closed convex subset of
$\mathfrak{T}( \H)$ consisting of positive operators with unit trace
called \emph{states} \cite{H-SCI,N&Ch}. Denote by $\B_+(\H)$ the
cone of positive operators in $\B(\H)$.

We will use the Greek letters $\rho,\sigma,...$ for trace-class
operators (not only for states) to distinguish their from bounded
operators which will be denoted $A,B,...$

The support $\mathrm{supp}\rho$ of a positive operator $\rho$ is the
orthogonal complement to its kernel.

Denote by $I_{\mathcal{H}}$ and $\mathrm{Id}_{\mathcal{H}}$ the unit
operator in a Hilbert space $\mathcal{H}$ and the identity
transformation of the Banach space $\mathfrak{T}(\mathcal{H})$
correspondingly.

Let $H(\rho)$ and $H(\rho\hspace{1pt}\|\hspace{1pt}\sigma)$ be
respectively the von Neumann entropy of a state $\rho$ and the
quantum relative entropy of states $\rho$ and $\sigma$
\cite{H-SCI,N&Ch}.

A finite or countable collection of states $\{\rho_i\}$ with the
corresponding probability distribution $\{\pi_i\}$ is called
\emph{ensemble} and denoted $\{\pi_i,\rho_i\}$. Its  Holevo quantity is defined as follows
$$
\chi(\{\pi_i,\rho_i\})\doteq\sum_i\pi_i H(\rho_i\hspace{1pt}\|\hspace{1pt}\bar{\rho})=H(\bar{\rho})-\sum_i\pi_i H(\rho_i),
$$
where $\bar{\rho}\doteq\sum_i\pi_i\rho_i$ is the \emph{average
state} of this ensemble and the second formula is valid under the
condition $H(\bar{\rho})<+\infty$ \cite{H-SCI,N&Ch}.

A  completely positive trace preserving linear map
$\Phi:\mathfrak{T}(\mathcal{H}_A)\rightarrow\mathfrak{T}(\mathcal{H}_B)$
is called  \emph{quantum channel} \cite{H-SCI,N&Ch}. It has the
Kraus representation
\begin{equation}\label{Kraus-rep}
\Phi(\rho)=\sum_{k}V_{k}\rho V^{*}_{k},\quad
\rho\in\,\mathfrak{T}(\mathcal{H}_A),
\end{equation}
where $\{V_{k}\}$ is a set of linear operators from $\mathcal{H}_A$
into $\mathcal{H}_B$ such that $\sum_{k}\!V^{*}_{k}V_{k}=I_{\H_A}.$

We will use  unitary dilation of a quantum channel
\cite{Caruso,Caruso+,H-SCI}: for a  channel
$\Phi:\mathfrak{T}(\mathcal{H}_A)\rightarrow\mathfrak{T}(\mathcal{H}_B)$
one can find separable Hilbert spaces $\H_D,\H_E$ for which
$\H_A\otimes\H_D\subseteq\H_B\otimes\H_E=\H$, a state $\rho_D$ in
$\S(\H_D)$ and an unitary operator  $U$  in the space $\H$ such that
this channel can be represented as follows
\begin{equation}\label{ch-rep}
\Phi(\rho)=\mathrm{Tr}_{\mathcal{H}_E}U\rho\otimes\rho_D U^{*},\quad
\rho\in\mathfrak{T}(\mathcal{H}_A).
\end{equation}
The quantum  channel
\begin{equation}\label{c-channel}
\widehat{\Phi}^w(\rho)=\mathrm{Tr}_{\mathcal{H}_B}U\rho\otimes\rho_D
U^{*},\quad \rho\in\mathfrak{T}(\mathcal{H}_A),
\end{equation}
is called \emph{weak complementary} to the channel $\Phi$
\cite{Caruso++}, \cite[Ch.6]{H-SCI}. If the state $\rho_D$ is pure
then (\ref{ch-rep}) is the Stinespring representation of the channel
$\Phi$ and the channel defined by (\ref{c-channel}) coincides with
the \emph{complementary} channel $\widehat{\Phi}$  to the channel
$\Phi$ \cite{D&Sh}. The weak complementary channel is not uniquely
defined (it depends on representation (\ref{ch-rep})), but the
complementary channel is unique: if
$\widehat{\Phi}':\mathfrak{T}(\mathcal{H}_A)\rightarrow\mathfrak{T}(\mathcal{H}_{E'})$
is a channel defined by (\ref{c-channel}) via another representation
(\ref{ch-rep}) (with pure state $\rho'_D$) then the channels
$\widehat{\Phi}$ and $\widehat{\Phi}'$ are isometrically equivalent
in the sense of the following definition \cite[the Appendix]{H-c-c}.

\begin{definition}\label{isom-eq}
Channels
$\Phi:\mathfrak{T}(\mathcal{H}_A)\rightarrow\mathfrak{T}(\mathcal{H}_B)$
and
$\Phi':\mathfrak{T}(\mathcal{H}_{A})\rightarrow\mathfrak{T}(\mathcal{H}_B')$
are \emph{isometrically equivalent} if there exists a partial
isometry $W:\mathcal{H}_B\rightarrow\mathcal{H}_{B'}$ such that
\begin{equation*}%\label{c-isom}
\Phi'(\rho)=W\Phi(\rho)W^*,\quad\Phi(\rho)=W^*\Phi'(\rho)W,\quad
\rho\in \T(\H_A).
\end{equation*}
\end{definition}

The notion of isometrical equivalence is very close to the notion of
unitary equivalence (see the remark after Definition 2 in
\cite{Sh-RC}).

\begin{definition}\label{c-q-def}
A channel $\Phi:\T(\H_A)\rightarrow\T(\H_B)$ is called
\emph{classical-quantum of discrete type} (briefly, \emph{discrete
c-q channel}) if it has the following representation
\begin{equation}\label{c-q-rep}
\Phi(\rho)=\sum_{i=1}^{\dim\H_A}\langle
i|\rho|i\rangle\sigma_i,\quad \rho\in \T(\H_A),
\end{equation}
where $\{|i\rangle\}$ is an orthonormal basis in $\H_A$ and
$\{\sigma_i\}$ is a collection of states in $\S(\H_B)$.
\end{definition}

We use the term "discrete" in this definition, since in infinite
dimensions there exist channels naturally called classical-quantum
which have no representation  (\ref{c-q-rep}), for example, Bosonic
Gaussian c-q channels \cite[Ch.12]{H-SCI}.

Discrete c-q channel (\ref{c-q-rep}), for which
$\,\sigma_i=\sigma\,$ for all $\,i$, is a \emph{completely
depolarizing} channel: $\Phi(\rho)=\sigma\Tr\rho$.

Following \cite{J-rev,O&Co} introduce the basic notion of this paper.
\begin{definition}\label{rev-def}
A channel $\Phi:\T(\H_A)\rightarrow\T(\H_B)$ is \emph{reversible}
with respect to a family $\S\subseteq\S(\H_A)$ if there exists a
channel $\,\Psi:\T(\H_B)\rightarrow\T(\H_A)$ such that
$\,\rho=\Psi\circ\Phi(\rho)\,$ for all $\,\rho\in\S$.
\end{definition}

In \cite{J&P,P-sqc} this property is called sufficiency of the
channel $\Phi$ for the family $\S$.

We will call $\Psi$ and  $\S$ in Def. \ref{rev-def} \emph{reversing
channel} and \emph{reversed family} respectively.

\begin{definition}\label{comp}
A family $\S$ of states in $\S(\H)$ is \emph{complete} if for any
nonzero positive operator $A$ in $\B(\H)$ there exists a state
$\rho\in\S$ such that $\Tr A\rho>0$.
\end{definition}

A family $\{|\varphi_{\lambda}\rangle\langle\varphi_{\lambda}|\}_{\lambda\in\Lambda}$ of pure states in
$\S(\H)$ is complete if and only if the linear hull of the family
$\{|\varphi_{\lambda}\rangle\}_{\lambda\in\Lambda}$ is dense in $\H$.

By separability of $\H$ an arbitrary complete family of states in
$\S(\H)$ contains a countable complete subfamily \cite[Lemma
2]{J&P}.

\section{General conditions for reversibility}

Now we consider general conditions for reversibility of a  channel
with respect to arbitrary families of states (pure states for the
most part).

We begin with the following observation showing the "face property"
of a set of all channels reversible with respect to a given family
of states.

\begin{property}\label{face} \emph{Let $\,\Phi_1$ and $\,\Phi_2$  be
 quantum channels from $\T(\H_A)$ to $\T(\H_B)$ and
$\,\Phi=p\,\Phi_1+(1-p)\Phi_2$, where $p\in(0,1)$. If the  channel
$\,\Phi$ is reversible with respect to a family $\,\S$ of states in
$\,\S(\H_A)$ then the channels $\,\Phi_1$ and $\,\Phi_2$ are
reversible with respect to the family $\,\S$.}
\end{property}

\textbf{Proof.} By Definition \ref{rev-def} reversibility of a
channel with respect to a given family of states is equivalent to
its reversibility with respect to any dense countable subfamily of
this family. So, since the space $\T(\H_A)$ is separable, we may
assume that the family $\S$ is countable.

Let $\S=\{\rho_i\}$ and $\{\pi_i\}$ be a nondegenerate probability
distribution with finite Shannon entropy. Then the Holevo quantity
of the ensemble $\{\pi_i, \rho_i\}$ is finite. Let
$\bar{\rho}=\sum_i\pi_i\rho_i$ be the average state of this
ensemble. By reversibility of the channel $\Phi$ with respect to the
family $\S$ we have
$$
\begin{array}{ll}
\!\!\!\displaystyle\sum_i\pi_i  H(\rho_i\hspace{1pt}\|\hspace{1pt}\bar{\rho})&
\!\!\!\displaystyle=\sum_i\pi_i H(\Phi(\rho_i)\hspace{1pt}\|\hspace{1pt}\Phi(\bar{\rho}))\\ &
\!\!\!\displaystyle=\sum_i\pi_i
H(p\,\Phi_1(\rho_i)+(1-p)\Phi_2(\rho_i)\hspace{1pt}\|\hspace{1pt}p\,\Phi_1(\bar{\rho})+(1-p)\Phi_2(\bar{\rho}))
\\ & \!\!\!\displaystyle \leq p\sum_i\pi_i
H(\Phi_1(\rho_i)\hspace{1pt}\|\hspace{1pt}\Phi_1(\bar{\rho}))+(1-p)\sum_i\pi_i
H(\Phi_2(\rho_i)\hspace{1pt}\|\hspace{1pt}\Phi_2(\bar{\rho}))\\ & \displaystyle \!\!\!\leq
p\sum_i\pi_i H(\rho_i\hspace{1pt}\|\hspace{1pt}\bar{\rho})+(1-p)\sum_i\pi_i
H(\rho_i\hspace{1pt}\|\hspace{1pt}\bar{\rho})=\sum_i\pi_i H(\rho_i\hspace{1pt}\|\hspace{1pt}\bar{\rho}),
\end{array}
$$
where the inequalities follow from monotonicity and joint convexity
of the relative entropy. Thus equalities hold in both these
inequalities and hence the channels $\Phi_1$ and $\Phi_2$ preserve
the Holevo quantity of the ensemble $\{\pi_i, \rho_i\}$. By Theorem
2 in \cite{J&P} (with the Remark after it) this implies
reversibility the channels $\Phi_1$ and $\Phi_2$ with respect to the
family $\S$. $\square$

Note that the assertion of Proposition \ref{face} is not inverted:
reversibility of channels $\Phi_1$ and $\Phi_2$ with respect to some
family $\S$ does not imply reversibility of their convex mixture
with respect to this family. The simplest example is given by
unitary channels $\Phi_1$ and $\Phi_2$.

Petz's theorem implies the following necessary condition for
reversibility of a quantum channel with respect to complete families
of states (for families of orthogonal states this condition is also
sufficient for reversibility).

\begin{theorem}\label{main}
\emph{Let $\,\S=\{\rho_i\}$ be a complete family of states in
$\,\S(\H_A)$, $\,\{\pi_i\}$ a nondegenerate probability distribution
and
$\,\{A_i=\pi_i[\hspace{1pt}\bar{\rho}\hspace{1pt}]^{-1/2}\rho_i[\hspace{1pt}\bar{\rho}\hspace{1pt}]^{-1/2}\}$
- the corresponding resolution of the identity in $\H_A$, where
$\bar{\rho}=\sum_i\pi_i\rho_i$.}

\emph{If a channel $\,\Phi:\T(\H_A)\rightarrow\T(\H_B)$ is
reversible with respect to the family $\,\S$ then any its weak
complementary channel $\,\widehat{\Phi}^w$ has the Kraus
representation}
\begin{equation}\label{s-k-r}
\widehat{\Phi}^w(\rho)=\sum_{i,j}W_{ij}\rho
W_{ij}^*\quad\textit{such that}\quad A_i=\sum_{j}W_{ij}^*W_{ij}
\quad\textit{for all}\,\; i.
\end{equation}
\emph{It follows, in particular, that}
\begin{equation}\label{rank-ineq}
\max_{i,j}\mathrm{rank} W_{ij}\leq \max_{i}\mathrm{rank}
\rho_{i}\quad\textit{and}\quad\min_i\max_j\mathrm{rank} W_{ij}\leq
\min_{i}\mathrm{rank} \rho_{i}.
\end{equation}

\emph{If
$\,\mathrm{supp}\hspace{1pt}\rho_i\perp\mathrm{supp}\hspace{1pt}\rho_k$
for all $i\neq k$ then existence of  Kraus representation
(\ref{s-k-r}) for the channel $\,\widehat{\Phi}^w=\widehat{\Phi}$ is
equivalent to reversibility of the channel $\,\Phi$ with respect to
the family $\,\S$}.
\end{theorem}

\textbf{Proof.} The Kraus representation (\ref{s-k-r}) for the
complementary channel $\widehat{\Phi}$ is constructed in the proof
of Theorem 3 in \cite{Sh-RC}.

Let $\widehat{\Phi}^w$ be a weak complementary channel to the
channel $\Phi$ defined by formula (\ref{c-channel}) and
$\rho_D=\sum_{k}\lambda_k\rho^k_D$ a pure states decomposition of
the state $\rho_D$. Then $\Phi=\sum_{k}\lambda_k\Phi_k$ and
$\widehat{\Phi}^w=\sum_{k}\lambda_k\widehat{\Phi}_k^w$, where
$\Phi_k$ and $\widehat{\Phi}_k^w$ are channels defined by formulae
(\ref{ch-rep}) and (\ref{c-channel}) with $\rho^k_D$ instead of
$\rho_D$. Since the state $\rho^k_D$ is pure, we have
$\widehat{\Phi}_k^{w}=\widehat{\Phi}_k$ for each $k$.

By Proposition \ref{face} reversibility of the channel $\Phi$ with
respect to the family $\S$ implies reversibility of all the channels
$\Phi_k$ with respect to this family. Thus, as mentioned before, all
the channels $\widehat{\Phi}_k^{w}=\widehat{\Phi}_k$ have Kraus
representation (\ref{s-k-r}). Hence the same property holds for the
channel $\widehat{\Phi}^w=\sum_{k}\lambda_k\widehat{\Phi}_k^w$.

If
$\,\mathrm{supp}\hspace{1pt}\rho_i\perp\mathrm{supp}\hspace{1pt}\rho_k$
for all $i\neq k$ then $A_i$ is the projector onto
$\,\mathrm{supp}\hspace{1pt}\rho_i$ for each $i$. Representation
(\ref{s-k-r}) of the channel $\widehat{\Phi}$ implies
$\widehat{\widehat{\Phi}}(\rho)=\sum_{i,j,k,l}\Tr[W_{ij}\rho
W^*_{kl}]|i\otimes j\rangle\langle k\otimes l|$ (cf. \cite{H-c-c})
and hence
$\,\mathrm{supp}\hspace{1pt}\widehat{\widehat{\Phi}}(\rho_i)\perp\mathrm{supp}\hspace{1pt}\widehat{\widehat{\Phi}}(\rho_k)$
for all $i\neq k$. It follows, since  $\Phi$ and
$\widehat{\widehat{\Phi}}$ are isometrically equivalent channels,
that the channel $\Phi$ is reversible with respect to the family
$\{\rho_i\}$ (by Lemma 1 in \cite{Sh-RC}). $\square$

In analysis of reversibility of a channel with respect to
noncomplete families of pure states we will need the following
notion.
\begin{definition}\label{subch}
The restriction of a channel $\Phi:\T(\H_A)\rightarrow\T(\H_B)$ to
the subspace $\T(\H_0)$ of $\T(\H_A)$, where $\H_0$ is a nontrivial
subspace of $\H_A$, is called \emph{subchannel of $\Phi$
corresponding to the subspace $\H_0$} and is denoted
$\Phi|_{\T(\H_0)}$.
\end{definition}

The (weak) complementary channel to the subchannel of a channel
$\Phi$ corresponding to any subspace $\H_0\subset\H_A$ coincides
with the subchannel of the (weak) complementary channel
$\widehat{\Phi}$ corresponding to the subspace $\H_0$, i.e.
\begin{equation}\label{comp-sub}
\widehat{\Psi}=\widehat{\Phi}|_{\T(\H_0)}\quad\text{and}\quad
\widehat{\Psi}^w=\widehat{\Phi}^w|_{\T(\H_0)},\quad\text{where}\quad
\Psi=\Phi|_{\T(\H_0)}.
\end{equation}

\begin{remark}\label{main-n-c-f}
It follows from (\ref{comp-sub}) that the reversibility conditions
in Theorem \ref{main} are generalized to noncomplete family $\S$ by
replacing the channels $\widehat{\Phi}^w$ and $\widehat{\Phi}$ by
their subchannels corresponding to the subspace
$\,\H_A^{\S}=\bigvee_{\rho\in\S}\mathrm{supp}\hspace{1pt}\rho$.
\end{remark}

\begin{remark}\label{main-r} The first relation in (\ref{rank-ineq})
shows that reversibility of a channel $\Phi$ with respect to a
complete family of states of rank $\leq r$ implies that any its weak
complementary channel $\widehat{\Phi}^w$ is $\,r$-partially
entanglement-breaking \cite{p-e-b-ch}. Thus, by Theorem \ref{main}
and Remark \ref{main-n-c-f}, to prove that a channel $\Phi$ is not
reversible with respect to any family $\S$ of states of rank $\leq
r$ it suffices to find its weak complementary channel
$\widehat{\Phi}^w$ and a state $\omega$ in $\S(\H^{\S}_A\otimes\K)$
such that
\begin{equation*}
\text{either}\quad
SN(\widehat{\Phi}^w\otimes\id_{\K}(\omega))>r\quad \text{or}\quad
E(\widehat{\Phi}^w\otimes\id_{\K}(\omega))>\log r,
\end{equation*}
where $SN$ is the Schmidt number and $E$ is any convex entanglement
monotone coinciding on the set of pure states with the entropy of a
partial state, in particular, $E=EoF$ \cite{P&V}.
\end{remark}

The second relation in (\ref{rank-ineq}) can be used to show
nonreversibility of a channel $\Phi$ with respect to families
containing at least one finite rank state. In particular, for
complete families it suffices to find a weak complementary channel
$\,\widehat{\Phi}^w$ such that any its Kraus representation
(\ref{Kraus-rep}) consists of infinite rank operators $V_k$. We will
use this way in the proof of Corollary \ref{new-c} in Section IV.2.

Theorem \ref{main} gives the following criteria for reversibility
with respect to orthogonal families.
\begin{corollary}\label{main-c}
\emph{Let $\,\Phi:\T(\H_A)\rightarrow\T(\H_B)$ be a quantum channel
and $\,\S=\{\rho_i\}$ a family of mutually orthogonal states in
$\,\S(\H_A)$. Let $P_i$ be the projector on the support of the state
$\rho_i$ for each $i$ and
$\,\H_A^{\S}=\bigoplus_i\mathrm{supp}\hspace{1pt}\rho_i$. The
following statements are equivalent:}
\begin{enumerate}[(i)]
  \item \emph{the channel $\,\Phi$ is reversible with respect to the family $\,\S$;}
  \item \emph{the subchannel $\Phi|_{\T(\H_A^{\S})}$ is isometrically equivalent to the channel
$$
\T(\H_A^{\S})\ni\rho\mapsto\Psi(\rho)=\sum_{i,j,k,l}\Tr[W_{ij}\rho
W^*_{kl}]|i\otimes j\rangle\langle k\otimes l|,
$$
where $\{W_{ij}\}$ is a set of operators such that
$\sum_{j}W_{ij}^*W_{ij}=P_i$ for all $\,i$;}
  \item \emph{$\widehat{\Phi}(\rho)=\sum_{i,j}W_{ij}\rho
W_{ij}^*$ for all $\rho\in\T(\H^{\S}_A)$, where $\{W_{ij}\}$ is the
same set as in $\mathrm{(ii)}$;}
  \item \emph{$\,P_i\widehat{\Phi}^*(A)P_k=0$ for all $A\in\B(\H_E)$ and all $\,i\neq k$;}
  \item \emph{$\,\{P_i\}\subset P\hspace{1pt}\Phi^*(\B_{+}(\H_B))P$, where $P=\sum_iP_i$ is the projector onto $\H_A^{\S}$.}

\end{enumerate}
\emph{If $\,\mathrm{(i)}$ holds then $\,\mathrm{(iii)}$ and
$\,\mathrm{(iv)}$ are valid for any weak complementary channel
$\widehat{\Phi}^w$ to the channel $\,\Phi$.}
\end{corollary}

If the family $\,\S$ is complete (i.e. $\H_A^{\S}=\H_A$) then
$\,\mathrm{(ii)}$ gives a description (up to isometrical
equivalence) of the set of all quantum channels reversible with
respect to $\S$.

\textbf{Proof.} By passing to the subchannel of $\Phi$ corresponding
to the subspace $\H_A^{\S}$ we may consider that $\S$ is a complete
family.

$\mathrm{(i)\Leftrightarrow(iii)}$ follows from Theorem \ref{main}
and Remark \ref{main-n-c-f}. $\mathrm{(ii)\Leftrightarrow(iii)}$
follows from the standard representation of a complementary channel
\cite[formula (11)]{H-c-c}. $\mathrm{(iii)\Rightarrow(iv)}$ is
easily verified.

$\mathrm{(iv)\Rightarrow(i)}$ For given $i$ and $k\neq i$ it follows
from $\mathrm{(iv)}$ that
$\widehat{\Phi}(|\varphi\rangle\langle\psi|)=0$ for any vectors
$\varphi\in \mathrm{supp}\hspace{1pt}\rho_i$ and $\psi\in
\mathrm{supp}\hspace{1pt}\rho_k$. By the definition of a
complementary channel this implies
$\mathrm{supp}\Phi(|\varphi\rangle\langle\varphi|)\perp\mathrm{supp}\Phi(|\psi\rangle\langle\psi|)$.
It follows that
$\mathrm{supp}\Phi(\rho_i)\perp\mathrm{supp}\Phi(\rho_k)$ for all
$i\neq k$ and hence the channel $\Phi$ is reversible with respect to
the family $\{\rho_i\}$.

$\mathrm{(ii)\Rightarrow(v)}$ Since $\Psi^*(A)=\sum_{i,j,k,l}\langle
k\otimes l|A|i\otimes j\rangle W^*_{kl}W_{ij}$, we have
$P_i=\sum_{j}W_{ij}^*W_{ij}=\Psi^*(|i\rangle\langle i|\otimes
I_{\H^{\{|j\rangle\}}})$, where $\H^{\{|j\rangle\}}$ is the Hilbert
space with the basis $\{|j\rangle\}$.

$\mathrm{(v)\Rightarrow(iii)}$ follows from the proof of Theorem 3
in \cite{Sh-RC}.

The last assertion of the corollary follows from Theorem \ref{main}
and Remark \ref{main-n-c-f}. $\square$

Now we consider conditions for reversibility of a quantum  channel
with respect to arbitrary families of pure states.

By Lemma 5 in \cite{Sh-RC} any  family $\S$ of pure states in
$\S(\H)$ has the unique (finite or countable) decomposition
\begin{equation}\label{OND-d}
\S=\bigcup_{k=1}^n\S_k\quad
(n\leq\dim\textstyle\bigvee_{\rho\in\S}\mathrm{supp}\hspace{1pt}\rho),
\end{equation}
where $\{\S_k\}_{k=1}^n$ is a collection of  orthogonally
non-decomposable (OND) families (this means that there is no
subspace $\H_0$ such that some states (not all) from $\S_k$ lie in
$\H_0$, while the others -- in $\H^{\perp}_0$) mutually orthogonal
in the sense that $\rho\perp\sigma$ if $\rho\in \S_k$ and $\sigma\in
\S_l$, $k\neq l$.

The following theorem is an extended and strengthened version of
Theorem 4 in \cite{Sh-RC}.

\begin{theorem}\label{grc}
\emph{Let $\,\Phi:\T(\H_A)\rightarrow\T(\H_B)$ be a quantum channel
and $\,\S$ a family of pure states in $\,\S(\H_A)$ with
decomposition (\ref{OND-d}) into OND subfamilies. Let $
\,\H_A^{\S}=\bigvee_{\rho\in\S}\mathrm{supp}\hspace{1pt}\rho$,$\;\H^{\S}_B=\bigvee_{\rho\in\S}\mathrm{supp}\hspace{1pt}\Phi(\rho)$,
$$
\,m=\min\left\{\,\dim\left[\,\ker
P_{\S}\Phi^*(\cdot)P_{\S}\,\cap\,\B(\H^{\S}_B)\,\right]+1,\,
\dim\H^{\S}_B\,\right\}
$$
where $P_{\S}$ is the projector on the subspace $\H_A^{\S}$, and
$\,\{P_k\}_{k=1}^n$ the orthogonal resolution of the identity in
$\H_A^{\S}$ corresponding to decomposition (\ref{OND-d}). The
following statements are equivalent:}
\begin{enumerate}[(i)]
  \item \emph{the channel $\,\Phi$ is reversible with respect to the family $\,\S$;}
  \item \emph{the channel $\,\Phi$ is reversible with respect to the family}
  $$
  \hat{\S}=\left\{\rho\in\S(\H^{\S}_A)\,\left|\;\rho=\sum_{k=1}^n P_k\rho P_k\right.\right\};
  $$\vspace{-15pt}
  \item \emph{the subchannel $\widehat{\Phi}|_{\T(\H_A^{\S})}$ is a discrete c-q channel having the representation
\begin{equation}\label{s-rep}
\,\widehat{\Phi}(\rho)=\displaystyle\sum_{k=1}^n[\Tr
P_k\rho]\sigma_k,\quad \rho\in\T(\H_A^{\S}),
\end{equation}
where $\{\sigma_k\}$ is a set of states in $\S(\H_E)$ such that
$\,\mathrm{rank}\hspace{1pt}\sigma_k\!\leq m$ for all $\,k;$}

\item \emph{the subchannel $\Phi|_{\T(\H_A^{\S})}$ is isometrically equivalent to the
channel}
$$
\Psi(\rho)=\sum_{k,l=1}^nP_k\rho
P_l\otimes\sum_{p,t=1}^m\langle\psi_t^l|\psi_p^k\rangle
|p\rangle\langle t|\vspace{-5pt}
$$
\emph{from $\,\T(\H^{\S}_A)$ into $\,\T(\H^{\S}_A\otimes\H_m)$,
where $\{|\psi_{p}^k\rangle\}$ is a collection of vectors  in a
separable Hilbert space such that $\,\sum_{p=1}^m\|\psi_{p}^k\|^2=1$
and $\langle\psi_{t}^k|\psi_{p}^k\rangle=0$ for all $\,p\neq t$ for
each
 $k$ and $\,\{|p\rangle\}_{p=1}^m$ is an orthonormal basis in $\,\H_m$.}
\end{enumerate}

\emph{If $\,\widehat{\Phi}^w$ is a weak complementary channel to the
channel $\,\Phi$ defined by (\ref{c-channel}) via the state $\rho_D$
then $\hspace{1pt}\mathrm{(i)}$ implies that
$\,\widehat{\Phi}^w|_{\T(\H_A^{\S})}$ is a discrete c-q channel
having representation (\ref{s-rep}) in which $\{\sigma_k\}$ is a set
of states in $\,\S(\H_E)$ such that
$\;\mathrm{rank}\hspace{1pt}\sigma_k\!\leq
\dim\H^{\S}_B\times\mathrm{rank}\hspace{1pt}\rho_D\,$ for all
$\,k$.}
\end{theorem}

\textbf{Proof.} The first assertion of the theorem follows from
Theorem 4 in \cite{Sh-RC} applied to the subchannel of $\Phi$
corresponding to the subspace $\H_A^{\S}$ and (\ref{comp-sub}).

Let $\widehat{\Phi}^w$ be a weak complementary channel to the
channel $\Phi$ defined by formula (\ref{c-channel}) and
$\rho_D=\sum_{i=1}^r\lambda_i\rho^i_D$ be a pure states
decomposition of the state $\rho_D$, where
$r=\mathrm{rank}\hspace{1pt}\rho_D$. Then
$\Phi=\sum_{i=1}^r\lambda_i\Phi_i$ and
$\widehat{\Phi}^w=\sum_{i=1}^r\lambda_i\widehat{\Phi}_i^w$, where
$\Phi_i$ and $\widehat{\Phi}_i^w$ are channels defined by formulae
(\ref{ch-rep}) and (\ref{c-channel}) with $\rho^i_D$ instead of
$\rho_D$. Since the state $\rho^i_D$ is pure, we have
$\widehat{\Phi}_i^{w}=\widehat{\Phi}_i$ for each $i$.

By Proposition \ref{face} reversibility of the channel $\Phi$ with
respect to the family $\S$ implies reversibility of the channels
$\Phi_i$, $i=\overline{1,r}$, with respect to this family. By the
first assertion of the theorem
$\widehat{\Phi}_i^w(\rho)=\widehat{\Phi}_i(\rho)=\sum_{k=1}^n[\Tr
P_k\rho]\sigma^i_k$ for all $\rho\in\S(\H_A^{\S})$, where
$\{\sigma^i_k\}$ is a set of states in $\S(\H_E)$ such that
$\mathrm{rank}\sigma^i_k\leq \dim\H^{\S}_B$, for each $i$. Hence
$$
\widehat{\Phi}^w(\rho)=\sum_{i=1}^r\lambda_i\widehat{\Phi}_i^w(\rho)=\sum_{k=1}^n[\Tr
P_k\rho]\sum_{i=1}^r\lambda_i\sigma^i_k,\quad \rho\in\T(\H_A^{\S}).\quad \square
$$

\begin{remark}\label{grc-r} If the family $\S$ in Theorem \ref{grc} is nonorthogonal then the collection $\{P_k\}_{k=1}^n$ contains
at least one projector $P_{k_0}$ of rank $>1$. By the implication
$\mathrm{(i)\Rightarrow(ii)}$ in Theorem \ref{grc} reversibility of
the channel $\Phi$ with respect to this family $\S$ implies its
reversibility with respect to the family of all states supported by
the subspace $\H_{k_0}=P_{k_0}(\H_A)$, i.e. its \emph{perfect
reversibility}  on the subspace $\H_{k_0}$ in terms of
\cite[Ch.10]{H-SCI}. Theorem \ref{grc} also shows that reversibility
of the channel $\Phi$ with respect to the family $\S$ implies that
the subchannels $\widehat{\Phi}|_{\T(\H_{k_0})}$ and
$\widehat{\Phi}^w|_{\T(\H_{k_0})}$ are completely depolarizing.
\end{remark}

Theorem \ref{grc} (with Remark \ref{grc-r}) shows that analysis of
reversibility properties of a quantum channel requires conditions
for existence of discrete c-q subchannels and of completely
depolarizing subchannels of the (weak) complementary channel. The
following lemma gives such conditions expressed in terms of the
kernel (null set) of a channel.

\begin{lemma}\label{ref-l}
\emph{Let $\,\Psi:\T(\H_A)\rightarrow\T(\H_B)$ be a quantum
channel.}

A) \emph{The channel $\,\Psi$ has no discrete c-q subchannels  if
and only if the set $\,\ker\Psi$ does not contain 1-rank operators.}

B) \emph{The channel $\,\Psi$ has discrete c-q subchannels but it
has no completely depolarizing subchannels if and only if the set
$\,\ker\Psi$ contains 1-rank operators but it does not contain the
operators
\begin{equation}\label{operators}
|\varphi\rangle\langle\psi|\quad\textit{and}\quad
|\varphi\rangle\langle\varphi|-|\psi\rangle\langle\psi|
\end{equation}
simultaneously for all unit vectors $\varphi$ and $\psi$ in
$\H_A$.}\footnote{Since $\Psi$ is a trace-preserving map, any 1-rank
operator in $\ker\Psi$ has the form $|\varphi\rangle\langle\psi|$,
where $\varphi$ and $\psi$ are orthogonal vectors in $\H_A$.}

C) \emph{The channel $\,\Psi$ has completely depolarizing
subchannels if and only if  the set $\,\ker\Psi$ contains operators
(\ref{operators}) for some unit vectors $\varphi$ and $\psi$ in
$\H_A$.}
\end{lemma}

\textbf{Proof.} The assertions of the lemma follow from Lemma
\ref{c-q-cond-l} in Appendix A. $\square $

To describe reversibility properties of a  channel $\Phi$ with
respect to families of pure states it is convenient to introduce the
\emph{reversibility index}
$\ri(\Phi)=[\,\ri_1(\Phi),\ri_2(\Phi)\,]$, in which the both
components take the values $0,1,2$. The first component
$\ri_1(\Phi)$ characterizes reversibility of the channel $\Phi$ with
respect to (w.r.t.) complete families of pure states as follows
\begin{description}
    \item [$\ri_1(\Phi)=0\,$] if $\,\Phi\,$  is not reversible w.r.t. any
    complete family $\S$ of pure
    states;
    \item [$\ri_1(\Phi)=1\,$] if $\,\Phi\,$  is reversible w.r.t. a complete orthogonal family $\S$ of pure
    states but it is not reversible w.r.t. any
    complete nonorthogonal family $\S$ of pure
    states;
    \item [$\ri_1(\Phi)=2\,$] if $\,\Phi\,$ is reversible w.r.t. a complete nonorthogonal family
    $\S$ of pure
    states.
\end{description}

The value of $\ri_1(\Phi)$ can be interpreted geometrically as
follows: $\ri_1(\Phi)>0$ means existence of an orthonormal basis
$\{|\varphi_i\rangle\}$ of the space $\H_A$ such that
\begin{equation}\label{vn-e}
    \mathrm{supp}\hspace{1pt}\Phi(|\varphi_i\rangle\langle\varphi_i|)\perp\mathrm{supp}\hspace{1pt}\Phi(|\varphi_j\rangle\langle\varphi_j|)\qquad
    \forall\, i\neq j,
\end{equation}
if the channel $\Phi$ is perfectly reversible on a subspace spanned
by some vectors of this basis  then $\ri_1(\Phi)=2$, otherwise
$\ri_1(\Phi)=1$ (this follows from Remark \ref{grc-r}).

The second component $\ri_2(\Phi)$ characterizes reversibility of
the channel  $\Phi$ with respect to noncomplete families of pure
states and is defined similarly to $\ri_1(\Phi)$ with the "complete
family $\S$" replaced by "noncomplete family $\S$".

So that $\,\ri(\Phi)=01\,$ means that the channel $\Phi$  is not
reversible with respect to any family of pure states which is either
complete or nonorthogonal, but it is reversible with respect to some
noncomplete orthogonal family.

By Remark \ref{grc-r} the value of $\ri_2(\Phi)$ has the clear
geometrical interpretation: $\ri_2(\Phi)=2$ means existence of a
subspace of $\H_A$ on which the channel $\Phi$ is perfectly
reversible, if there are no such subspaces but there exists an
orthonormal set $\{|\varphi_i\rangle\}$ of vectors in $\H_A$ such
that (\ref{vn-e}) holds then $\ri_2(\Phi)=1$. This implies the
following observation.

\begin{remark}\label{zec}
If $\Phi$ is a finite dimensional channel then $\ri_2(\Phi)$
characterizes positivity of one-shot zero-error capacities of $\Phi$
as follows:
$$
\ri_2(\Phi)=0 \;\Leftrightarrow\; \bar{C}_0(\Phi)=0,\qquad
\ri_2(\Phi)=2 \;\Leftrightarrow\; \bar{Q}_0(\Phi)>0,
$$
(so, $\ri_2(\Phi)=1$ means that $\bar{C}_0(\Phi)>0$ but
$\bar{Q}_0(\Phi)=0$), where $\bar{C}_0(\Phi)$ and $\bar{Q}_0(\Phi)$
are the one-shot zero-error classical and quantum capacities of the
channel $\Phi$ respectively \cite{ZEC,W&Co}.
\end{remark}

It follows from the definition that the reversibility index can take
the values
$$
00,\quad 01,\quad 02,\quad 11,\quad 12,\quad 22.
$$

By using the below Corollary \ref{grc-c+} it is easy to construct a
channel with any reversibility index from the above list excepting
the index $12$. Existence of a channel $\Phi$ with
$\,\ri(\Phi)=12\,$ is an interesting open question.\footnote{It
seems intuitively that any discrete c-q channel (\ref{c-q-rep}) with
$\,\sigma_i\neq\sigma_j$ for all $\,i\neq j$ can not have completely
depolarizing subchannels, but I can not find a formal proof. I would
be grateful for any comments.}

\begin{corollary}\label{grc-c+}
\emph{Let $\Phi$ be a quantum channel and $\widehat{\Phi}$ its
complementary channel. Then
$$
\begin{array}{l}
\{\,\ri(\Phi)=00\,\}\,\Leftrightarrow\{\,\widehat{\Phi} \textit{ satisfies condition A of Lemma \ref{ref-l}}\,\},\\
\{\,\ri(\Phi)=01\,\}\,\Leftrightarrow\{\,\widehat{\Phi} \textit{ is
not discrete c-q and satisfies condition B of Lemma \ref{ref-l}}\,\},\\
\{\,\ri(\Phi)=02\,\}\,\Leftrightarrow\{\,\widehat{\Phi} \textit{ is
not discrete c-q and
satisfies condition C of Lemma \ref{ref-l}}\,\},\\
\{\,\ri(\Phi)=11\,\}\,\Leftrightarrow\{\,\widehat{\Phi} \textit{ is
discrete C-Q and satisfies condition B of Lemma \ref{ref-l}}\,\}, \\
\{\,\ri(\Phi)=12\,\}\,\Leftrightarrow\{\,\widehat{\Phi} \textit{ is
discrete C-Q and satisfies condition C of Lemma \ref{ref-l}}\,\},\\
\{\,\ri(\Phi)=22\,\}\,\Leftrightarrow\{\,\widehat{\Phi} \textit{ is
discrete c-q channel (\ref{c-q-rep}) with $\,\sigma_i=\sigma_j$ for
some $\,i\neq j$}\,\},
\end{array}
$$
where "discrete C-Q" denotes discrete c-q channel (\ref{c-q-rep})
with $\,\sigma_i\neq\sigma_j$ for all $\,i\neq j$.}

\emph{For a weak complementary channel $\,\widehat{\Phi}^w$ to the
channel $\,\Phi$ the following implications hold~\footnote{Here and
in what follows $X_1X_2\leq Y_1Y_2$ means that $X_1\leq Y_1$ and
$X_2\leq Y_2$.}
$$
\begin{array}{l}
\{\,\ri(\Phi)= 00\,\}\Leftarrow\,\{\,\widehat{\Phi}^w \textit{
satisfies condition A of Lemma
\ref{ref-l}}\;\},\\
\{\,\ri(\Phi)\geq02\,\}\,\Rightarrow\{\,\widehat{\Phi}^w \textit{
satisfies condition C of Lemma
\ref{ref-l}}\;\},\\
\{\,\ri(\Phi)\geq11\,\}\,\Rightarrow\{\,\widehat{\Phi}^w \textit{ is
a discrete c-q channel}\;\}.
\end{array}
$$}
\end{corollary}
\textbf{Proof.} All the above assertions follow from Theorem
\ref{grc}, Remark \ref{grc-r} and Lemma \ref{ref-l}. $\square$

We will show in the next section  that the reversibility index takes
the values $00,01,02$ and $22$ on the class of Bosonic linear
channels.

\section{Reversibility of Bosonic linear channels}

Let $\mathcal{H}_{X}$ $(X=A,B,...)$ be the space of irreducible representation of
the Canonical Commutation Relations (CCR)
\begin{equation*}
W_X(z)W_X(z^{\prime })=\exp
\left(-\textstyle{\frac{\mathrm{i}}{2}}\,\Delta_{X}(z,z^{\prime})\right)
W_X(z^{\prime }+z),\quad z,z'\in Z_X,
\end{equation*}
where  $(Z_{X},\Delta _{X})$ is a symplectic space and $W_{X}(z)$
are the Weyl operators \cite{Caruso,E&W},\cite[Ch.12]{H-SCI}. We
will also use the symbol $\Delta _{X}$ for the matrix of the form
$\Delta _{X}$, i.e.
$\Delta_{X}(z,z^{\prime})=z^{\top}\Delta_{X}z^{\prime}$. Denote by
$s_X$ the number of modes of the system $X$, i.e. $2s_X=\dim
Z_X$.\footnote{Some basic notions concerning symplectic spaces are
presented in Appendix B.}

A Bosonic linear channel  $\Phi_{K,f}
:\mathfrak{T}(\mathcal{H}_{A})\rightarrow
\mathfrak{T}(\mathcal{H}_{B})$ is defined via the action of its dual
$\Phi_{K,f}^{\ast }:\mathfrak{B}(\mathcal{H}_{B})\rightarrow
\mathfrak{B}(\mathcal{H}_{A})$ on the Weyl operators:
\begin{equation}
\Phi_{K,f}^{\ast}(W_{B}(z))=W_A(Kz)f(z),\quad z\in Z_B, \label{blc}
\end{equation}
where $K:Z_{B}\rightarrow Z_{A}$ is a linear operator, and $f(z)$ is
a complex continuous function on $Z_B$ such that $\,f(0)=1\,$ and
the matrix with the elements
$f(z_s-z_r)\exp\left(\,\frac{\mathrm{i}}{2}\,z_s^{\top}[\Delta_B-K^{\top}\Delta_A
K] z_r\right)$ is positive for any finite subset $\{z_s\}$ of $Z_B$
\cite{H-BC,H-EBC}. This channel is also called quasi-free
\cite{DVV}.

We will assume existence of a Bosonic  unitary dilation for the channel $\Phi_{K,f}$, i.e. existence of such Bosonic systems $D$ and $E$ that this channel can be represented as a restriction
of a corresponding unitary evolution of the composite system $AD=BE$ (described by the
symplectic space $Z=Z_A\oplus Z_D=Z_B\oplus Z_E$) provided that the system $D$ is in a particular state $\rho_D$. This means that
\begin{equation}\label{ud}
\Phi_{K,f}^*(W_B(z))=\Tr_{\H_D}(I_{\H_A}\!\otimes\rho_D)U_T^*(W_B(z)\otimes
I_{\H_E})U_T, \quad z\in Z_B,
\end{equation}
where $U_T$ is the unitary operator in the space
$\H_A\otimes\H_D\cong\H_B\otimes\H_E$ implementing the symplectic
transformation
\begin{equation}\label{t-m}
T=\left[\begin{array}{ll}
        K & L\\
        K_D & L_D
        \end{array}\right]
\end{equation}
of the space $Z$ (here $L:Z_E\rightarrow Z_A,\,K_D:Z_B\rightarrow Z_D,\,L_D:Z_E\rightarrow Z_D$ are appropriate linear operators) \cite{Caruso, Caruso+, H-SCI,H-EBC}. Note that
\begin{equation}\label{f-rep}
    f(z)=\phi_{\rho_D}(K_Dz),
\end{equation}
where $\phi_{\rho_D}$ is the characteristic function of the state
$\rho_D$.

The weak complementary channel (see Section II) is defined as
follows
\begin{equation}\label{w-c-ch}
\begin{array}{rl}
\displaystyle[\widehat{\Phi}_{K,f}^w]^*(W_E(z))\;=&\!\!\Tr_{\H_D}(I_{\H_A}\!\otimes\rho_D)U_T^*(I_{\H_B}\!\otimes
W_E(z))U_T\\
\displaystyle=&\!\!\Tr_{\H_D}(I_{\H_A}\!\otimes\rho_D)(W_A(Lz)\otimes
W_D(L_Dz))\\
\displaystyle=&\!\! W_A(Lz)\phi_{\rho_D}(L_Dz), \quad z\in Z_E.
\end{array}
\end{equation}
Thus $\widehat{\Phi}_{K,f}^w$ is a Bosonic linear channel as well.
If the state $\rho_D$ is pure then
$\widehat{\Phi}_{K,f}^w=\widehat{\Phi}_{K,f}$ is the complementary
channel to the channel $\Phi_{K,f}$.

\begin{remark}\label{ud-r}
Unitary dilation (\ref{ud})-(\ref{t-m}) does not exist for all
Bosonic linear channels (it suffices to note that (\ref{f-rep})
implies $|f(z)|=1\Leftrightarrow f(z)=1$), but one can conjecture
that any Bosonic linear channel can be transformed by appropriate
displacement unitaries to Bosonic linear channel for which such
dilation exists.\footnote{I would be grateful for any comments
concerning this question.} This conjecture is true for Bosonic
Gaussian channels (see Section IV.2).
\end{remark}

\subsection{Reversibility conditions}

We begin with the following observation concerning reversibility
with respect to complete families. \pagebreak

\begin{property}\label{nr-cf}
\emph{Let $\,\Phi_{K,f}$ be a noisy (not noiseless \cite{n-ch})
Bosonic linear channel for which  unitary dilation
(\ref{ud})-(\ref{t-m}) exists. The channel $\,\Phi_{K,f}$ is not
reversible with respect to a complete family $\,\S$ in the following
cases:
\begin{itemize}
    \item $\S$ consists of pure states;
%    \item $\S$ consists of mutually orthogonal states and contains at least one  finite rank state.
    \item $\S$ consists of orthogonal states at least one of which has finite rank.
\end{itemize}
The channel $\,\Phi_{K,f}$ is reversible with respect to a
particular complete family $\,\S$ of orthogonal infinite rank states
if and only if $\,Z_f\doteq\{z\in Z_B\,|\,f(z)=1\}\neq\{0\}$.}
\end{property}

\textbf{Proof.} It suffices to prove the first assertion in the
second case, since reversibility of a channel with respect to a
complete family of pure states implies its reversibility with
respect to some complete family of orthogonal pure states
\cite[Corollary 2]{Sh-RC}.

Let $\S=\{\rho_i\}$ be a complete family of orthogonal states and
$\{P_i\}$ the corresponding orthogonal resolution of the identity in
$\H_A$ ($P_i$ is the projector onto $\mathrm{supp}\rho_i$). Let
$\Phi_{L,g}=\widehat{\Phi}^w_{K,f}$ be a weak complementary channel
to the channel $\Phi$ defined by (\ref{w-c-ch}), where
$g(z)=\phi_{\rho_D}(L_Dz)$. Since $\Phi_{K,f}$ is not noiseless,
$\Phi_{L,g}$ is not completely depolarizing, i.e. $L\neq0$. By
Corollary \ref{main-c} ($\mathrm{(i)\Rightarrow(iv)}$) reversibility
of the channel $\Phi_{K,f}$ with respect to the family $\S$ implies
\begin{equation}\label{sr}
g(z)W_A(Lz)=g(z)\sum_{i}P_iW_A(Lz)P_i\quad \forall z\in Z_E.
\end{equation}

If $\mathrm{rank}\rho_{i_0}<+\infty$ for some $i_0$ then (\ref{sr})
implies that the operator $W_A(z_0)$, where $z_0$ is a nonzero
vector in $\mathrm{Ran}L$, commutes with the finite dimensional
projector $P_{i_0}$. This contradicts to the well known fact that
the Weyl operators have no purely point spectrum.

If $Z_f=\{0\}$ then, since (\ref{f-rep}) implies $Z_f=\ker K_D$,
Lemma \ref{main-l} below shows that $\mathrm{Ran}
L=[K(Z_f)]^{\perp}=Z_A$. It follows that the family
$\{W_A(Lz)\}_{z\in Z_E}$ acts irreducibly on $\H_A$ and hence
(\ref{sr}) can not be valid for any orthogonal resolution of the
identity $\{P_i\}$.

Let $Z_f=\ker K_D\neq\{0\}$. Consider the von Neumann algebras
$\mathcal{A}$ and $\mathcal{B}$ generated respectively by the
families $\{W_A(Kz)\}_{z\in Z_f}$ and $\{W_B(z)\}_{z\in Z_f}$. By
the second assertion of Lemma \ref{main-l} below the restriction of
the operator  $K$ to the subspace $Z_f$ is nondegenerate and
symplectic (i.e. $\Delta_A(Kz_1,Kz_2)=\Delta_B(z_1,z_2)$ for all
$z_1,z_2\in Z_f$). This implies that the restriction of the dual map
$\Phi^*_{K,f}$ to the algebra $\mathcal{B}\subseteq\B(\H_B)$ is a
$*$-isomorphism between the algebras $\mathcal{B}$ and $\mathcal{A}$
(see details in \cite[Sect.2]{Sh-SBC}). It follows that for any
orthogonal resolution of the identity $\{P_i\}$ in $\mathcal{A}$
there exists an unique orthogonal resolution of the identity
$\{Q_i\}$ in $\mathcal{B}$ such that $P_i=\Phi^*_{K,f}(Q_i)$ for all
$i$ and hence
\begin{equation}\label{inv-rel}
\Phi_{K,f}\left(\S(\H_A^i)\right)\subseteq\S(\H_B^i)\quad\forall i,
\end{equation}
where $\,\H_A^i=P_i(\H_A)$ and $\,\H_B^i=Q_i(\H_B)$ (so that
$\,\H_A=\bigoplus_i\H_A^i$ and $\,\H_B=\bigoplus_i\H_B^i$).

Let $\{\rho_i\}$ be a family of states in $\S(\H_A)$ such that
$\mathrm{supp}\rho_i=\H_A^i$ for all $i$. It follows from
(\ref{inv-rel}) that the channel $\,\Phi_{K,f}$ is reversible with
respect to the orthogonal family $\{\rho_i\}$ and that the simplest
reversing channel has the form
\begin{equation}\label{r-ch}
\Psi(\sigma)=\sum_i[\Tr Q_i\sigma]\rho_i,\quad
\sigma\in\S(\H_B).\;\;\square
\end{equation}

Now we consider reversibility of Bosonic linear channels with
respect to arbitrary families of pure states. For these channels the
reversibility index (introduced in Section III) can take the values
$\,00,\hspace{1pt}01,\hspace{1pt}02,\hspace{1pt}22\,$.

\begin{theorem}\label{lb-rc}
\emph{Let $\,\Phi_{K,f}$ be a Bosonic linear channel for which
unitary dilation (\ref{ud})-(\ref{t-m}) exists and
$\,Z_f\doteq\{z\in Z_B\,|\,f(z)=1\}\,$. Then $\ri(\Phi_{K,f})=22$ if
and only if $\Phi_{K,f}$ is a noiseless channel (see \cite{n-ch}).
Otherwise}\vspace{-5pt}
\begin{equation}\label{main-s}
\begin{array}{l}
\{\,\ri(\Phi_{K,f})=00\,\}\,\Leftrightarrow\,\{\,Z_f=\{0\}\,\},\\
\{\,\ri(\Phi_{K,f})=01\,\}\,\Leftrightarrow\,\{\,Z_f\,\textit{is a
nontrivial isotropic subspace of }\,Z_B\,\},\\
\{\,\ri(\Phi_{K,f})=02\,\}\,\Leftrightarrow\,\{\,\exists\;z_1,z_2\in
Z_f\,\;\textit{such that}\;\, \Delta_B(z_1,z_2)\neq0\,\}.
\end{array}
\end{equation}
\end{theorem}

A  description (in the Schrodinger representation) of reversed
families of pure states for the channel $\,\Phi_{K,f}$ in the cases
$\ri(\Phi_{K,f})=01,02$ is given in Section IV.3.

\textbf{Proof.} The first assertion of the theorem follows from
Proposition \ref{nr-cf}.

In proving the second one we may consider (by using a purification
procedure if necessary) that (\ref{ud})-(\ref{t-m}) is a Stinespring
dilation for the channel $\,\Phi_{K,f}$, i.e. the state $\rho_D$ is
pure. Then the complementary channel to the channel $\,\Phi_{K,f}$
is a Bosonic linear channel defined by (\ref{w-c-ch}).

Since (\ref{f-rep}) implies $Z_f=\ker K_D$, Lemma \ref{main-l} below
shows that $\mathrm{Ran} L=[K(Z_f)]^{\perp}$, $\ker K\cap Z_f=\{0\}$
and that $\Delta_A(Kz_1,Kz_2)=\Delta_B(z_1,z_2)$ for all $z_1,z_2\in
Z_f$. Hence all the statements in (\ref{main-s}) follow from
Corollary \ref{grc-c+} and Proposition \ref{c-q-cond-p} in Appendix
A. $\square$

\begin{remark}\label{lb-rc-c-r}
Sufficiency of the reversibility conditions (\ref{main-s}) can be
shown without using Corollary \ref{grc-c+} by explicit construction
of reversing channels for particular orthogonal and non-orthogonal
families of pure states.

Reversibility of the channel $\Phi_{K,f}$ with respect to some
orthogonal families of pure states under the condition
$Z_f\neq\{0\}$ can be shown by repeating the arguments from the
proof of Proposition \ref{nr-cf} and by taking the family
$\{\rho_i\}$ consisting of pure states such that
$\mathrm{supp}\rho_i\subseteq \H_A^i$ for all $i$. The simplest
reversing channel in this case is given by (\ref{r-ch}).

Consider now how to prove the implication $"\Leftarrow"$ in the
third statement in (\ref{main-s}). In this case one can construct
\emph{Bosonic linear} reversing channels for families of all states
supported by particular subspaces of $\H_A$.

Indeed, if $Z_B^0$ is a nontrivial symplectic subspace of $Z_f$ then
the second assertion of Lemma \ref{main-l} below shows that the
restriction $K^0$ of the operator $K$ to the subspace $Z_B^0$ is a
symplectic embedding of this subspace into $Z_A$. Let
$Z_A^0=K(Z_B^0)$. Then $Z_X=Z_X^0\oplus Z_X^*$, where
$Z_X^*=[Z_X^0]^{\perp}$, and hence $\H_X=\H_X^0\otimes\H_X^*$,
$X=A,B$. Let $\sigma$ be a given arbitrary state in $\S(\H_A^*)$ and
$\S_{\sigma}=\{\rho\otimes\sigma\,|\,\rho\in\S(\H_A^0)\}\subset\S(\H_A)$.
Let $\Psi^0(\cdot)=U_{K^0}(\cdot)U^*_{K^0}$ be a channel from
$\T(\H_B^0)$ to $\T(\H_A^0)$, where $U_{K^0}$ is the unitary
operator from $\H_B^0$ onto $\H_A^0$ implementing the symplectic
transformation $K^0:Z_B^0\rightarrow Z_A^0$, and $\Psi^*$ be the
completely depolarizing channel from $\T(\H_B^*)$ to $\T(\H_A^*)$
with the output state $\sigma$. Then $\Psi^0\otimes\Psi^*$ is a
Bosonic linear channel from $\T(\H_B)$ to $\T(\H_A)$ and it is easy
to see that
$\Psi^0\otimes\Psi^*(\Phi_{K,f}(\omega))=\omega\hspace{1pt}$ for all
$\hspace{1pt}\omega\in\S_{\sigma}$.
\end{remark}

\begin{lemma}\label{main-l}
\emph{Let $\,T:Z_B\oplus Z_E\rightarrow Z_A\oplus Z_D$ be a
symplectic transformation  defined by matrix (\ref{t-m}). Then
$[\mathrm{Ran} L]^{\perp}=K(\ker K_D)$ and $\ker
K_D=\Delta_BK^{\top}\!\Delta_A\left([\mathrm{Ran}
L]^{\perp}\right)$, where $[\mathrm{Ran} L]^{\perp}$ is the
skew-orthogonal complement to the subspace $\mathrm{Ran}
L=\{Lz\}_{z\in Z_E}\subseteq Z_A$ (see Appendix B).}\footnote{We
will always use this sense of the symbol $"\perp"$ dealing with a
subspace of a symplectic space.}

\emph{The restriction of the operator $K$ (correspondingly,
$\Delta_BK^{\top}\!\Delta_A$) to the subspace $\ker K_D$
(correspondingly, $[\mathrm{Ran} L]^{\perp}$) is nondegenerate and
symplectic, i.e. it preserves the corresponding skew-symmetric forms
$\Delta_X$, $X=A,B$.}
\end{lemma}

This lemma shows that for a given Bosonic linear channel
$\,\Phi_{K,f}$ the subspace $\mathrm{Ran} L$ is determined by the
set $Z_f=\ker K_D$ and does not depend on a choice of its unitary
dilation (\ref{ud})-(\ref{t-m}). It implies that the algebra
generated by the Weyl operators $W_A(z)$, $z\in K(Z_f)^{\perp}$,
coincides with the noncommutative graph of the channel
$\,\Phi_{K,f}$ (in terms of \cite{W&Co}).

\textbf{Proof.} Note first that $[\mathrm{Ran} L]^{\perp}=\ker[L^{\top}\!\Delta_A]$.

Since the matrix $T$ defined in (\ref{t-m}) is symplectic, we have (cf. \cite{H-EBC})
\begin{equation}\label{one-eq}
\begin{array}{rrl}
    \Delta_B\,=&\!\!K^{\top}\!\Delta_AK\;\,  + & \!\!K_D^{\top}\Delta_DK_D,\\
    0\;\,=&\!\!L^{\top}\!\Delta_AK\;\;\,  + & \!\!L_D^{\top}\Delta_DK_D,\\
    \Delta_E\,=&\!\!L^{\top}\!\Delta_AL\;\;\;  + & \!\!L_D^{\top}\Delta_DL_D.
\end{array}
\end{equation}
Since the group of symplectic matrices is closed under
transposition, the matrix $T^{\top}$ is symplectic and hence we have the
following equations (similar to (\ref{one-eq}))
\begin{equation}\label{two-eq}
\begin{array}{rrl}
    \Delta_A\,=& K\,\Delta_BK^{\top}\;\;\;\;\;  + & \!\!L\,\Delta_EL^{\top},\\
    0\;\,=&\!\!K_D\Delta_BK^{\top}\;\;\;  + & \!\!L_D\Delta_EL^{\top},\\
    \Delta_D\,=&\!\!K_D\Delta_BK_D^{\top}\;\;\;  + & \!\!L_D\Delta_EL_D^{\top}.
\end{array}
\end{equation}
The second equations in (\ref{one-eq}) and (\ref{two-eq}) imply
respectively
\begin{equation}\label{incl}
K(\ker K_D)\subseteq\ker [L^{\top}\!\Delta_A],\quad \Delta_BK^{\top}\!\Delta_A(\ker
[L^{\top}\!\Delta_A])\subseteq\ker K_D,
\end{equation}
while the first equations in (\ref{one-eq}) and (\ref{two-eq}) show
that
$$
\ker K\cap\ker K_D=\{0\},\quad\ker[\Delta_BK^{\top}\!\Delta_A]\cap\ker
[L^{\top}\!\Delta_A]=\{0\},
$$
since the matrices $\Delta_A$ and $\Delta_B$ are nondegenerate. It
follows, by the dimension arguments, that $"="$ holds in the both
inclusions in (\ref{incl}).

The last assertions of the lemma directly follow from  the first
equations in (\ref{one-eq}) and (\ref{two-eq}).

\begin{corollary}\label{lb-rc-c+} \emph{If $\,\det[\Delta_B-K^{\top}\!\Delta_A K]\neq0\,$ then
$\,\ri(\Phi_{K,f})=00$, i.e. the channel $\,\Phi_{K,f}$ is not reversible with respect to any families of pure states.}
\end{corollary}

\textbf{Proof.} It is shown in \cite{H-EBC} that the condition $\,\det[\Delta_B-K^{\top}\!\Delta_A K]\neq0\,$ implies existence
of unitary dilation (\ref{ud})-(\ref{t-m}) for the channel $\Phi_{K,f}$ in which $D=B$ and $K_D$ is a nondegenerate quadratic matrix. $\square$

\subsection{The case of Gaussian channels}

Bosonic Gaussian channels are Bosonic linear channels defined by
(\ref{blc}) with the function
$$
f(z)=\exp \left(\,
\mathrm{i}\hspace{1pt}l\hspace{1pt}z-\textstyle\frac{1}{2}\hspace{1pt}z^{\top
}\alpha \hspace{1pt}z\,\right),
$$
where $l\,$ is a $\,2s_B$-dimensional real row and $\,\alpha\,$ is a
real symmetric $\,(2s_B)\times(2s_B)$ matrix satisfying the
inequality $\alpha \geq \pm \frac{\mathrm{i}}{2}\left[ \Delta
_{B}-K^{\top }\Delta _{A}K\right]$ \cite{Caruso,E&W,H-SCI}.

Any such channel can be transformed by appropriate displacement
unitaries to the Bosonic Gaussian channel with $\,l=0\,$ and the
same matrix $\,\alpha\,$ for which unitary dilation
(\ref{ud})-(\ref{t-m}) always exists with Gaussian state $\rho_D$
\cite{Caruso,H-SCI}.  In this case $\alpha=K_D^{\top}\alpha_DK_D$,
where $\alpha_D$ is the covariance matrix of $\rho_D$. Thus all the
above results can be reformulated for Bosonic Gaussian channels by
noting that $\,Z_f=\ker K_D=\ker \alpha\,$ (since the matrix
$\alpha_D$ is nondegenerate). In particular, Theorem \ref{lb-rc} is
reformulated  as follows.

\begin{corollary}\label{lb-rc-c}
\emph{Let $\,\Phi$ be a noisy (not noiseless \cite{n-ch}) Bosonic
Gaussian channel with the parameters $K,\,l,\,\alpha$. Then}
\begin{equation*}\;\label{main-s+}
\begin{array}{l}
\{\,\ri(\Phi)=00\,\}\,\Leftrightarrow\,\{\,\det\alpha\neq 0\,\},\\
\{\,\ri(\Phi)=01\,\}\,\Leftrightarrow\,\{\,\ker\alpha\;\,\textit{is
a
nontrivial isotropic subspace of}\;\,Z_B\,\},\\
\{\,\ri(\Phi)=02\,\}\,\Leftrightarrow\,\{\,\exists\;z_1,z_2\in
\ker\alpha\,\;\textit{such that}\;\, \Delta_B(z_1,z_2)\neq0\,\}.
\end{array}
\end{equation*}
\end{corollary}

Physically, this characterization of reversibility of a Gaussian
channel $\Phi$ is intuitively clear, since in the Heisenberg picture
the condition $\det\alpha\neq 0$ means that the channel $\Phi^{*}$
injects quantum noise in all canonical variables of the system $B$,
while degeneracy of the matrix $\alpha$ is equivalent to existence
of noise-free canonical variables. Corollary \ref{lb-rc-c} shows
that
\begin{description}
  \item[-] the channel $\Phi$ is reversible with respect to some families of pure states if and only if the set of noise-free canonical variables is nonempty;
  \item[-] the channel $\Phi$ is reversible \emph{only} with respect to some orthogonal families of pure states if and only if all noise-free canonical variables commute;
  \item[-] the channel $\Phi$ is reversible with respect to some nonorthogonal families of
  pure states (and hence it is perfectly reversible on a particular subspace) if and only if there are noncommuting noise-free canonical variables.
\end{description}

\textbf{Example: one-mode Gaussian channels.} The simplest Bosonic
Gaussian channels are one-mode channels for which  $\dim Z_A=\dim
Z_B=2$.

A classification of all one-mode Gaussian channels is obtained in
\cite{H-1MGC}, where it is shown that there exist the following
canonical types
$$
A_1[N],\;\; A_2[N],\;\; B_1,\;\; B_2[N],\;\; C[k,N]\;
(k>0,k\neq1),\;\; D[k,N] \;(k>0)
$$
of such channels (the parameter $N\geq0$ denotes the level of
noise, see details in \cite{H-SCI,H-1MGC}).

By Corollary \ref{lb-rc-c} all one-mode Gaussian channels are not
reversible with respect to any families of pure states excepting the
noiseless channel $B_2[0]$ and the channel $B_1$ which has
reversibility index $01$. All reversed families of pure states for
the channel $B_1$ are described in Section IV.3.

By Proposition \ref{nr-cf} the channel $B_1$ is the only noisy
one-mode Gaussian channel reversible with respect to some complete
orthogonal families of (infinite rank) states. Applying the proof of
Proposition \ref{nr-cf} to the channel $B_1$ we have
$\mathcal{A}=\mathcal{B}=L_{\infty}(\mathbb{R})$ and hence any
orthogonal resolution of the identity $\{P_i\}$ in $\mathcal{A}$
corresponds to a decomposition $\{D_i\}$ of $\mathbb{R}$ into
disjoint measurable subsets. In this case $\{Q_i\}=\{P_i\}$ and
$\H_A^i=\H_B^i=L_{2}(D_i)$ is the subspace of
$\H_A=\H_B=L_{2}(\mathbb{R})$ consisting of functions supported by
$D_i$.

Thus, the subset $\S(L_{2}(D_i))\subset\S(L_{2}(\mathbb{R}))$ is
mapped by the channel $B_1$ into itself for each $i$. This
conclusion agrees with the explicit formula for the channel $B_1$
(formula (7.1) in \cite{Ivan} with $q$ replaced by $p$).

It follows that the channel $B_1$ is reversible with respect to any
family of states $\{\rho_i\}$ such that $\rho_i\in\S(L_{2}(D_i))$
for each $i$. One can expect that all reversed families for the
channel $B_1$ have such form. For families of pure states this is
proved in Section IV.3 (see the example).

In regard to reversibility of one-mode Gaussian channels with
respect to nonorthogonal families of mixed states we have the
following partial result.
\begin{corollary}\label{new-c}
\emph{Let $\Phi$ be a one-mode Gaussian channels of any type
excepting $\,B_1$, $B_2[0]$ and $\,C[k,0]$ with $k>1$. Then the
channel $\Phi$ is not reversible with respect to any complete family
of states containing at least one finite rank state.}
\end{corollary}

\textbf{Proof.} As shown in \cite{Ivan} all operators of any Kraus
representation of the channel $C[k,0]$ with $k\neq1$ have infinite
rank. By  Theorem \ref{main} (see Remark \ref{main-r}) this  implies
nonreversibility of the complementary channel to the channel
$C[k,0]$ with $k\neq1$ with respect to any complete family of states
containing at least one finite rank state. This implies
nonreversibility of the channels $C[k,0]$ with $k<1$ and $D[k,0]$
(complementary channels to one-mode Gaussian channels are described
in \cite{H-SCI,H-1MGC}). Nonreversibility of all the others channels
excepting the channels $B_1$, $B_2[0]$ and $C[k,0]$ with $k>1$ can
be shown by using their representation in the form $\Psi\circ\Phi$,
where $\Phi$ is either the channel $C[k,0]$  with $k<1$ or the
channel $D[k,0]$ (see Table I in \cite{Ivan}). $\square$

\subsection{Explicit forms of reversed families}

Now we will give an explicit description of reversed families of
pure states for the channel $\Phi_{K,f}$. We will show that these
families are completely determined by the subspace $K(Z_f)$ of
$Z_A$. By Theorem \ref{lb-rc} it suffices to consider the cases
$\ri(\Phi_{K,f})=01,02$.

\textbf{ri}$\mathbf{(\Phi_{K,f})=01}$. By Theorem \ref{lb-rc} and
Lemma \ref{main-l} in this case $K(Z_f)$ is a nontrivial isotropic
subspace of $Z_A$ and hence the subspace
$\mathrm{Ran}L=[K(Z_f)]^{\perp}$ contains a maximal isotropic
subspace of $Z_A$. By Lemma \ref{sg-2} in Appendix B there exists a
symplectic basis $\{\tilde{e}_k, \tilde{h}_k\}$ in $Z_A$ such that
$\{\tilde{e}_1,...,\tilde{e}_{s_A},
\tilde{h}_{d+1},...,\tilde{h}_{s_A}\}$ is a basis in
$\mathrm{Ran}L$, $0< d\leq s_A$. If we identify the space $\H_A$
with the space $L_2(\mathbb{R}^{s_A})$ of complex-valued functions
of $s_A$ variables (which will be denoted $\xi_1,...,\xi_{s_A}$) and
the Weyl operators $W_A(\tilde{e}_k)$ and $W_A(\tilde{h}_k)$ with
the operators
$$
\psi(\xi_1,...,\xi_{s_A})\mapsto e^{\mathrm{i}\xi_k}
\psi(\xi_1,...,\xi_{s_A})\;\;\,\text{and}\;\;
\psi(\xi_1,...,\xi_{s_A})\mapsto
\psi(\xi_1,...,\xi_{k}+1,...,\xi_{s_A})
$$
then Theorem \ref{grc}, Lemma \ref{c-q-cond-l} and the proof of
Proposition \ref{c-q-cond-p} in Appendix A show that \emph{all}
reversed families of pure states for the channel $\Phi_{K,f}$
correspond to families $\{\psi_i\}$ of functions in
$L_2(\mathbb{R}^{s_A})$ with unit norm satisfying the following
condition
\begin{equation}\label{supp-cond+}
    \psi_i \cdot S_{y_{d+1},...,y_{s_A}}\psi_j=0\,\;(\textup{in}\; L_2(\mathbb{R}^{s_A}))\quad
    \forall(y_{d+1},...,y_{s_A})\in\mathbb{R}^{s_A-d},\; \forall\, i\neq j,
\end{equation}
where $S_{y_{d+1},...,y_{s_A}}$ is a shift operator
by the vector $(0,...,0,y_{d+1},...,y_{s_A})$:
$$
(S_{y_{d+1},...,y_{s_A}}\psi)(\xi_1,...,\xi_{s_A})=\psi(\xi_1,...,\xi_d,\xi_{d+1}+y_{d+1},...,\xi_{s_A}+y_{s_A}).
$$
This condition means, roughly speaking, that all shifts in $\mathbb{R}^{s_A}$ of the supports of the functions of
the family $\{\psi_i\}$ along the last $\,s_A-d\,$ coordinates do not intersect each other.

As an example of a reversed family one can take the family of
product pure states
$|\phi_i\otimes\varphi\rangle\langle\phi_i\otimes\varphi|$
corresponding to the family of functions
\begin{equation*}%\label{t-p-f}
\psi_i(\xi_1,...,\xi_{s_A})=\phi_i(\xi_1,...,\xi_{d})\varphi(\xi_{d+1},...,\xi_{s_A}),
\end{equation*}
where $\{\phi_i\}$ is a family of functions in $L_2(\mathbb{R}^d)$
with mutually disjoint supports and $\varphi$ is a given function in
$L_2(\mathbb{R}^{s_A-d})$.

\textbf{Example: one-mode Gaussian channel $B_1$.} In this case
$s_A=s_B=1$,
$$
K=\left[\begin{array}{cc}
        \;1\;&\; 0\; \\
        \;0\;&\;1
        \end{array}\right],\quad
\alpha=\left[\begin{array}{cc}
        \;0\;&\; 0\; \\
        \;0\;&\; 1/4
       \end{array}\right],\quad
Z_f=\ker\alpha=\left[\begin{array}{c}
        \;1\; \\
        \;0\;
        \end{array}\right].
$$
Hence $\tilde{e}_1=[1,0]^{\top}$, $\tilde{h}_1=[0,1]^{\top}$ (since
$K(Z_f)^{\perp}=K(Z_f)=\{\lambda\tilde{e}_1\}$) and condition
(\ref{supp-cond+}) shows that \emph{all} reversed families of pure
states for this channel have the form
$$
\{|\psi_i\rangle\langle \psi_i|\},\;\,
\textup{where}\;\{\psi_i\}\subset L_2(\mathbb{R})\,\; \textup{such
that}\;\, \psi_i\cdot\psi_j=0\,\; (\textup{in}\;
L_2(\mathbb{R}))\;\,\forall\, i\neq j.
$$
i.e., roughly speaking,  all reversed families of pure states
correspond to families of functions with mutually disjoint supports
(in agreement with the observations in Section IV.2 which show
sufficiency of this condition).

\textbf{ri}$\mathbf{(\Phi_{K,f})=02}$. By Theorem \ref{lb-rc}  in
this case there exists  a symplectic subspace $Z_B^0$ of $Z_f$. By
Lemma \ref{main-l} $Z_A^0=K(Z_B^0)$ is a symplectic subspace of
$[\mathrm{Ran}L]^{\perp}=K(Z_f)$. Let
 $\{\tilde{e}_k, \tilde{h}_k\}_{k=1}^{s_A}$ be a symplectic basis in $Z_A$ such that
$\{\tilde{e}_k, \tilde{h}_k\}_{k=1}^{d}$ is a symplectic basis in
$Z_A^0$. If we identify the space $\H_A$ with the space
$L_2(\mathbb{R}^{s_A})$ as before then Theorem \ref{grc}, Lemma
\ref{c-q-cond-l} and the proof of Proposition \ref{c-q-cond-p} in
Appendix A show that the channel $\Phi_{K,f}$ is perfectly
reversible on the subspaces
$L_2(\mathbb{R}^d)\otimes\{c|\varphi\rangle\},\, \varphi\in
L_2(\mathbb{R}^{s_A-d})$ (in agreement with the second part of
Remark \ref{lb-rc-c-r}).

\section{Applications}

By Petz's theorem reversibility properties of a quantum channel are
closely related to the question of
(non)\nobreakdash-\hspace{0pt}preserving the Holevo quantity of
arbitrary (discrete or continuous) ensembles of states under action
of this channel, i.e. to the question of validity of an equality in
the general inequality
\begin{equation}\label{chi-m}
    \chi(\Phi(\mu))\leq\chi(\mu),
\end{equation}
which holds, by monotonicity of the relative entropy, for any
channel $\Phi:\T(\H_A)\rightarrow\T(\H_B)$ and any generalized
ensemble $\mu$ of states in $\S(\H_A)$ (defined as a Borel
probability measure on $\S(\H_A)$, see \cite[Section 5]{Sh-RC}).

Denote by $\P(\S(\H_A))$ the set of all generalized ensembles of
pure states (probability measures on $\S(\H_A)$ supported by pure
states). Denote by $\P_\mathrm{c}(\S(\H_A))$ and
$\P_\mathrm{o}(\S(\H_A))$ the subsets of $\P(\S(\H_A))$ consisting
respectively of all ensembles with nondegenerate average state
(barycenter) and of all ensembles of mutually orthogonal pure
states.

For a given channel $\Phi:\T(\H_A)\rightarrow\T(\H_B)$ let
$\P(\Phi)$ be the subset of $\P(\S(\H_A))$ consisting of all
ensembles $\mu$ for which an equality holds in (\ref{chi-m}). The
version of Petz's theorem for continuous ensembles (Proposition 3 in
\cite{Sh-RC}) shows that:
$$
\begin{array}{l}
\{\,\ri(\Phi)=00\,\}\,\Rightarrow\,\{\,\P(\Phi)=\emptyset\,\},  \\
\{\,\ri(\Phi)=01\,\}\,\Rightarrow\,\{\,\P(\Phi)\subset\P_\mathrm{o}\backslash\P_\mathrm{c}\},  \\
\{\,\ri(\Phi)=02\,\}\,\Rightarrow\,\{\,\P(\Phi)\subset\P\backslash\P_\mathrm{c}\,\},  \\
\{\,\ri(\Phi)=11\,\}\,\Rightarrow\,\{\,\P(\Phi)\subset\P_\mathrm{o}\,\},  \\
\{\,\ri(\Phi)=12\,\}\,\Rightarrow\,\{\,\P(\Phi)\subset\P_\mathrm{o}\cup[\P\backslash\P_\mathrm{c}]\,\},
\end{array}
$$
where we write $\P_*$ instead of $\P_*(\S(\H_A))$ for brevity.

Thus, Corollary \ref{lb-rc-c} implies the following assertions.

\begin{corollary}\label{s-d-p}
\emph{Let $\,\Phi$ be a Gaussian channel with the
parameters $K,\,l,\,\alpha$.}

A) \emph{If $\,\Phi$ is not a noiseless channel then
\begin{equation}\label{chi-ineq}
\chi(\Phi(\mu))<\chi(\mu)
\end{equation}
for any ensemble $\mu$ of pure states with nondegenerate average
state;}

B) \emph{If $\;\ker\alpha$ is an isotropic subspace of $\,Z_B$ then
(\ref{chi-ineq}) holds for any nonorthogonal (in particular,
continuous) ensemble $\mu$ of pure states;}

C) \emph{If $\;\det\alpha\neq0$ then (\ref{chi-ineq}) holds
for any ensemble $\mu$ of pure states.}
\end{corollary}

By using the observations in Section IV.3 one can describe all
ensembles $\mu$ of pure states for which
$\chi(\Phi(\mu))=\chi(\mu)$. All such ensembles are completely
determined by the subspace $K(\ker\alpha)$.

Some applications of conditions for an equality (strict inequality)
in (\ref{chi-m}) to study of capacities of quantum channels are
considered in \cite[Section 5]{Sh-RC}.

\section*{Acknowledgments}

I am grateful to A.S.Holevo and to the participants of his seminar
"Quantum probability, statistic, information" (the Steklov
Mathematical Institute) for useful discussion. I am also grateful to
the referee for suggestions improving the paper. The work is
partially supported by the RAS research program and by RFBR grants
12-01-00319a and 13-01-00295a.

\section*{Appendix A: On discrete c-q subchannels and completely depolarizing subchannels of  Bosonic linear channels}

Note first that any nontrivial Bosonic linear channel $\Phi_{K,f}$
is not a discrete c-q channel (it is a discrete c-q channel if and
only if it is completely depolarizing). This immediately follows
from Definition \ref{c-q-def} and (\ref{blc}), since the Weyl
operator $W_A(Kz)$ has purely point spectrum for any $z\in Z_B$ if
and only if $K=0$.

In this section we explore necessary and sufficient conditions for
existence of discrete c-q subchannels and of completely depolarizing
subchannels of a Bosonic linear channel $\Phi_{K,f}$.

The following lemma shows that all discrete c-q subchannels and all
completely depolarizing subchannels of a quantum channel are
determined by its kernel (null set).

\begin{lemma}\label{c-q-cond-l}
\emph{Let $\,\Psi:\T(\H_A)\rightarrow\T(\H_B)$ be a quantum channel
and $\,\Pi(\Psi)$ the set of all families $\,\{|\psi_i\rangle\}$ of
unit vectors in $\H_A$ such that
$\Psi(|\psi_i\rangle\langle\psi_j|)=0$ for all $\,i\neq j$.}
\emph{The channel $\,\Psi$ has a discrete c-q subchannel
corresponding to a subspace $\H_0$, i.e.
\begin{equation*}
 \Psi(\rho)=\sum_i\langle \psi_i|\rho|\psi_i\rangle\sigma_i\quad
\forall\rho\in\T(\H_0),
\end{equation*}
where $\{|\psi_i\rangle\}$ is an orthonormal basis in $\H_0$, if and
only if $\,\{|\psi_i\rangle\}\in\Pi(\Psi)$.  Under this condition}
$$
\sigma_i=\sigma_j\quad
\Leftrightarrow\quad\Psi(|\psi_i\rangle\langle
\psi_i|-|\psi_j\rangle\langle \psi_j|)=0.
$$
\end{lemma}

\textbf{Proof.} It suffices to note that $\rho=\sum_{i,j}\langle
\psi_i|\rho|\psi_j\rangle|\psi_i\rangle\langle \psi_j|$ for any
$\rho\in\T(\H_0)$. $\square$

\begin{remark}\label{dual-c} The conditions
$\Psi(|\psi_i\rangle\langle\psi_j|)=0$ and
$\Psi(|\psi_i\rangle\langle \psi_i|)=\Psi(|\psi_j\rangle\langle
\psi_j|)$ can be expressed respectively as follows
$$
\langle \psi_i|\Psi^*(B)|\psi_j\rangle=0\;\;\;\forall
B\in\B(\H_B),\quad\langle \psi_i|\Psi^*(B)|\psi_i\rangle=\langle
\psi_j|\Psi^*(B)|\psi_j\rangle\;\;\;\forall B\in\B(\H_B),
$$
where $\Psi^*:\B(\H_B)\rightarrow\B(\H_A)$ is a dual map to the
channel $\Psi$.

If $\Psi=\Phi_{K,f}$ then, since the family $\{W_B(z)\}_{z\in Z_B}$
generates $\B(\H_B)$ and $f$ is a continuous function such that
$f(0)=1$, the above conditions can be rewritten as
\begin{equation*}
\langle \psi_i|W_A(Kz)|\psi_j\rangle=0\quad \forall z\in Z_B,
\eqno{(\textup{A}1)}
\end{equation*}
\begin{equation*}
\langle \psi_i|W_A(Kz)|\psi_i\rangle=\langle
\psi_j|W_A(Kz)|\psi_j\rangle\quad \forall z\in
Z_B.\eqno{(\textup{A}2)}
\end{equation*}
\end{remark}

By using Remark \ref{dual-c} it is easy to show that the set
$\Pi(\Phi_{K,f})$ introduced in Lemma \ref{c-q-cond-l} is empty if
and only if $\,\mathrm{rank}K=\dim Z_A$ and to describe all families
belonging to this set in the case $\,\mathrm{rank}K<\dim Z_A$. This
will be done in the proof of the following proposition.

\begin{property}\label{c-q-cond-p} \emph{The channel $\,\Phi_{K,f}$ has discrete c-q subchannels if and only if $\,\mathrm{rank}K<\dim Z_A$.
Under this condition all these subchannels are not completely
depolarizing if and only if
\begin{equation*}%\label{K-cond}
    \mathrm{Ran} K\doteq\{Kz\}_{z\in Z_B}\;\textit{contains a maximal isotropic subspace
    of }\,Z_A,\eqno{(\textup{A}3)}
\end{equation*}
which means that the subspace $[\mathrm{Ran} K]^{\perp}$ is
isotropic, i.e. there exist no $z_1,z_2\in\ker[K^{\top}\!\Delta_A]$
such that $\Delta_A(z_1,z_2)\neq0$}.
\end{property}

\textbf{Proof.} If $\mathrm{rank}K=\dim Z_A$ then the family
$\{W_A(Kz)\}_{z\in Z_B}$ of Weyl operators acts irreducibly on
$\H_A$. Hence condition (A-1) can not be valid.

If $\mathrm{rank}K<\dim Z_A$ and condition (A-3) holds then Lemma
\ref{sg-2} in Appendix B implies existence of a symplectic basis
$\{\tilde{e}_k, \tilde{h}_k\}$ in $Z_A$ such that
$\{\tilde{e}_1,...,\tilde{e}_{s_A},
\tilde{h}_{d+1},...,\tilde{h}_{s_A}\}$ is a basis in
$\mathrm{Ran}K$, $d\leq s_A$. Let $Z_B^0$ be a subspace of $Z_B$
with the basis $\{z^e_1,...,z^e_{s_A},z^h_{d+1},...,z^h_{s_A} \}$
such that $\tilde{e}_k=Kz^e_k$ for all $k=\overline{1,{s_A}}$ and
$\tilde{h}_k=Kz^h_k$ for all $k=\overline{d+1,{s_A}}$. Thus for any
vector $z\in Z_B^0$ represented as $z=\sum_{k=1}^{s_A} x_k
z^e_k+\sum_{k=d+1}^{s_A} y_k z^h_k$,
$(x_1,...,x_{s_A})\in\mathbb{R}^{s_A}$,
$(y_{d+1},...,y_{s_A})\in\mathbb{R}^{s_A-d}$ we have
$$
\begin{array}{c}
   \displaystyle W_A(Kz)=W_A\left(\sum_{k=1}^{s_A} x_k
Kz^e_k+\sum_{k=d+1}^{s_A} y_k
Kz^h_k\right)=W_A\left(\sum_{k=1}^{s_A} x_k
\tilde{e}_k+\sum_{k=d+1}^{s_A} y_k \tilde{h}_k\right)\\
   \displaystyle =\lambda W_A(x_1
\tilde{e}_1)\cdot...\cdot W_A(x_{s_A} \tilde{e}_{s_A})\cdot
W_A(y_{d+1} \tilde{h}_{d+1})\cdot...\cdot W_A(y_{s_A}
\tilde{h}_{s_A}),
 \end{array}
$$
where $\lambda=e^{\mathrm{i}
[x_{d+1}y_{d+1}+...+x_{s_A}y_{s_A}]}\neq 0$.

By identifying the space $\H_A$ with the space
$L_2(\mathbb{R}^{s_A})$ of complex-valued functions of $s_A$
variables (which will be denoted $\xi_1,...,\xi_{s_A}$) and the Weyl
operators $W_A(\tilde{e}_k)$ and $W_A(\tilde{h}_k)$ with the
operators
$$
\psi(\xi_1,...,\xi_{s_A})\mapsto e^{\mathrm{i}\xi_k}
\psi(\xi_1,...,\xi_{s_A})\;\;\,\text{and}\;\;
\psi(\xi_1,...,\xi_{s_A})\mapsto
\psi(\xi_1,...,\xi_{k}+1,...,\xi_{s_A})
$$
the equality in (A-1) for the vector $z$ can be rewritten as follows
\begin{equation*}%\label{c-q-sub+++}
\int \overline{\psi_i(\xi_1,...,\xi_{s_A})}
(S_{y_{d+1},...,y_{s_A}}\psi_j)(\xi_1,...,\xi_{s_A}) e^{\mathrm{i} (x_1\xi_1+...+x_{s_A}\xi_{s_A})}d\xi_1,...,d\xi_{s_A}=0,
\end{equation*}
where
$(S_{y_{d+1},...,y_{s_A}}\psi_j)(\xi_1,...,\xi_{s_A})=\psi_j(\xi_1,...,\xi_d,\xi_{d+1}+y_{d+1},...,\xi_{s_A}+y_{s_A})$.

This equality is valid for all
$(x_1,...,x_{s_A})\in\mathbb{R}^{s_A}$ and
$(y_{d+1},...,y_{s_A})\in\mathbb{R}^{s_A-d}$ (that is for all $z\in
Z_B^0$) if and only if
$$
\overline{\psi_i(\xi_1,...,\xi_{s_A})}
(S_{y_{d+1},...,y_{s_A}}\psi_j)(\xi_1,...,\xi_{s_A})=0
$$
for almost all $(\xi_1,...,\xi_{s_A})\in\mathbb{R}^{s_A}$ and all
$(y_{d+1},...,y_{s_A})\in\mathbb{R}^{s_A-d}$. Since
$\mathrm{Ran}K=K(Z_B^0)$, it implies that the set $\Pi(\Phi_{K,f})$
introduced in Lemma \ref{c-q-cond-l} consists of families
$\{\psi_i\}$ satisfying the condition
\begin{equation*}%\label{supp-cond}
    \psi_i \cdot S_{y_{d+1},...,y_{s_A}}\psi_j=0\,\;(\textup{in}\; L_2(\mathbb{R}^{s_A}))\quad
    \forall(y_{d+1},...,y_{s_A})\in\mathbb{R}^{s_A-d},\; \forall \,i\neq j.\eqno{(\textup{A}4)}
\end{equation*}
This condition means, roughly speaking, that all shifts in
$\mathbb{R}^{s_A}$ of the supports of the functions of the family
$\{\psi_i\}$ along the last $\,s_A-d\,$ coordinates do not intersect
each other.

As an example of a family satisfying condition (A-4) one can take
the family of functions
\begin{equation*}%\label{t-p-f+}
\psi_i(\xi_1,...,\xi_{s_A})=\phi_i(\xi_1,...,\xi_{d})\varphi(\xi_{d+1},...,\xi_{s_A}),\eqno{(\textup{A}5)}
\end{equation*}
where $\{\phi_i\}$ is a family of functions in $L_2(\mathbb{R}^d)$
with mutually disjoint supports and $\varphi$ is a given function in
$L_2(\mathbb{R}^{s_A-d})$. It is clear that this family (consisting
of tensor product vectors $|\phi_i\otimes\varphi\rangle$) is not a
"general solution" of (A-4).

To show that for any $i\neq j$ the equality in (A-2) can not be
valid for all $z\in Z_B$ note that this equality for the vector
$z=\sum_{k=1}^{s_A} x_kz^e_k\in Z_B^0$ can be rewritten as follows
\begin{equation*}%\label{c-q-sub+++}
\begin{array}{c}
\displaystyle\int |\psi_i(\xi_1,...,\xi_{s_A})|^2 e^{\mathrm{i}
(x_1\xi_1+...+x_{s_A}\xi_{s_A})}d\xi_1,...,d\xi_{s_A}\\\\=\displaystyle\int
|\psi_j(\xi_1,...,\xi_{s_A})|^2 e^{\mathrm{i}
(x_1\xi_1+...+x_{s_A}\xi_{s_A})}d\xi_1,...,d\xi_{s_A}.
\end{array}
\end{equation*}
Validity of this equality for all
$(x_1,...,x_{s_A})\in\mathbb{R}^{s_A}$ means that  the classical
characteristic functions of the probability densities $|\psi_i|^2$
and $|\psi_j|^2$ coincide. But this obviously contradicts to
condition (A-4).

If condition (A-3) is not valid then the subspace
$[\mathrm{Ran}K]^{\perp}$ contains a symplectic subspace $Z_A^0$.
Let
 $\{\tilde{e}_k, \tilde{h}_k\}_{k=1}^{s_A}$ be a symplectic basis in $Z_A$ such that
$\{\tilde{e}_k, \tilde{h}_k\}_{k=1}^{d}$ is a symplectic basis in
$Z_A^0$.

By identifying the space $\H_A$ with the space
$L_2(\mathbb{R}^{s_A})$ as before we see that the equalities in
(A-1) and in (A-2) can be rewritten respectively as
\begin{equation*}%\label{c-q-sub++}
\int \overline{\psi_i(\xi_1,...,\xi_{s_A})}
(W_A(Kz)\psi_j)(\xi_1,...,\xi_{s_A})d\xi_1,...,d\xi_{s_A}=0\eqno{(\textup{A}6)}
\end{equation*}
and
\begin{equation*}%\label{c-q-sub+++}
\begin{array}{cc}
\displaystyle\int \overline{\psi_i(\xi_1,...,\xi_{s_A})}
(W_A(Kz)\psi_i)(\xi_1,...,\xi_{s_A})d\xi_1,...,d\xi_{s_A}&\\\\=\displaystyle\int
\overline{\psi_j(\xi_1,...,\xi_{s_A})}
(W_A(Kz)\psi_j)(\xi_1,...,\xi_{s_A})d\xi_1,...,d\xi_{s_A},&
\end{array}\eqno{(\textup{A}7)}
\end{equation*}
where $\{\psi_i\}$ is an orthonormal family of functions in
$L_2(\mathbb{R}^{s_A})$.

Since for any $z\in Z_B $ the vector $Kz$ has no components
corresponding to the vectors $\tilde{e}_k,
\tilde{h}_k,\,k=\overline{1,d}$, one can satisfy equalities (A-6)
and (A-7) for all $i\neq j$ and all $z\in Z_B$ by taking the family
of functions (A-5), in which $\{\phi_i\}$ is an arbitrary
orthonormal basis in $L_2(\mathbb{R}^d)$ and $\varphi$ is a given
function in $L_2(\mathbb{R}^{s_A-d})$. Thus this family belongs to
the set $\Pi(\Phi_{K,f})$ introduced in Lemma \ref{c-q-cond-l} and
this lemma shows that the restriction of the channel $\Phi_{K,f}$ to
any subspace of the form
$L_2(\mathbb{R}^d)\otimes\{c|\varphi\rangle\}$, $\varphi\in
L_2(\mathbb{R}^{s_A-d})$, is completely depolarizing. $\square$

\section*{Appendix B: Some facts about symplectic spaces}

In what follows $Z$ is a $2s$-dimensional symplectic space with  the
nondegenerate skew-symmetric form $\Delta$ \cite{Arnold, H-SCI,
K&M}. The set of vectors $\{e_1,...,e_s, h_1,...,h_s\}$ is called
\emph{symplectic basis} in $Z$ if
$\Delta(e_k,e_l)=\Delta(h_k,h_l)=0$ for all $k,l$, but
$\Delta(e_k,h_l)=\delta_{kl}$. For an arbitrary subspace $L\subset
Z$ one can define its skew-orthogonal complement $L^{\perp}=\{z\in
Z\,|\,\Delta(z,z')=0\;\forall z'\in L\}$. Despite the fact that
$L\cap L^{\perp}\neq\{0\}$ in general, we always have the familiar
relations
\begin{equation*}
[L^{\perp}]^{\perp}=L\quad\textrm{and}\quad\dim L+\dim
L^{\perp}=\dim Z.
\end{equation*}

A linear transformation $T:Z\rightarrow Z$ is called
\emph{symplectic} if $\Delta(Tz_1,Tz_2)=\Delta(z_1,z_2)$ for all
$z_1,z_2\in Z$. A symplectic transformation maps any symplectic
basis to symplectic basis and vice versa: any two symplectic base
are related by the particular symplectic transformation.

A subspace $L$ of $Z$ is called \emph{symplectic} if the form
$\Delta$ is nondegenerate on $L$, in this case $L$ has even
dimension and can be considered as a symplectic space of itself. We
will use the following simple observation \cite{Arnold,K&M}.

\begin{lemma}\label{s-l} \emph{If $L$ is a symplectic subspace of $Z$ then
$L^{\perp}$ is a symplectic subspace of $Z$ and $Z=L+L^{\perp}$
(i.e. $Z=L\vee L^{\perp}$ and $L\cap L^{\perp}=\{0\}$).}
\end{lemma}

By joining  the symplectic base in $L$ and in $L^{\perp}$ we obtain
a symplectic basis in $Z$.

A subspace $L$ of $Z$ is called \emph{isotropic} if the form
$\Delta$ equals to zero on $L$. In this case $L$ has dimension $\leq
s$. We will use the following observation.

\begin{lemma}\label{i-l} \emph{If $L$ is an isotropic subspace of $Z$, $\dim L=d$, then
there exists a symplectic basis $\,\{\tilde{e}_k, \tilde{h}_k\}$ in
$Z$ such that $\,\{\tilde{e}_1,...,\tilde{e}_d\}$ is a basis in
$L$.}
\end{lemma}

\textbf{Proof.} Let $\{e_k, h_k\}$ be an arbitrary symplectic basis
in $Z$ and $L'$  the isotropic subspace of $Z$ generated by the
vectors $e_1,...,e_d$. Since the isotropic subspaces $L$ and $L'$
have the same dimension, there is a symplectic transformation $T$
such that $L=T(L')$ \cite{K&M}. The basis $\{\tilde{e}_k=Te_k,
\tilde{h}_k=Th_k\}$ has the required properties. $\square$

Now we can prove the lemma used in the proof of Proposition
\ref{c-q-cond-p}.

\begin{lemma}\label{sg-2}
\emph{Let $L$ be an arbitrary subspace of $Z$. Then there exists a symplectic
basis in $Z$ such that $\,\dim L\,$ vectors of this basis lie in
$L$.}
\end{lemma}

\textbf{Proof.} If the subspace $L$ is either symplectic or
isotropic then the assertion of the lemma follows respectively from
Lemma \ref{s-l} (with the remark after it)  and Lemma \ref{i-l}.

If the subspace $L$ is neither symplectic nor isotropic then
$$
L_1=L\cap L^{\perp}=\{z\in L\,|\,\Delta(z,z')=0,\,\forall z'\in
L\}
$$
is a nontrivial subspace of $L$. Let $L_2$ be an arbitrary subspace
such that $L=L_1+L_2$, i.e. $L=L_1\vee L_2$ and $L_1\cap L_2=\{0\}$.
Then the subspace $L_2$ is symplectic. Indeed, if there is a vector
$z_0\in L_2$ such that $\Delta(z_0,z)=0$ for all $z\in L_2$ then
$\Delta(z_0,z+z')=0$ for all $z'\in L_1$, $z\in L_2$, which implies
$z_0\in L_1$ and hence $z_0=0$.

By Lemma \ref{s-l} the subspace $L_2^{\perp}$ is symplectic. It
obviously contains the isotropic subspace $L_1$. By Lemma \ref{i-l}
there exists a symplectic basis $\{e_k, h_k\}$ in $L_2^{\perp}$ such
that $\{e_1,...,e_d\}$ is a basis in $L_1$. By joining this basis
and any symplectic basis in $L_2$ we obtain a basis with the
required properties. $\square$


\begin{thebibliography}{99}%{Литература}

\bibitem{Arnold} Arnold V.I.,  \emph{Mathematical Methods Of Classical Mechanics}  (Springer, 1989).

\bibitem{Caruso++} Caruso F., Giovannetti V. "Qubit quantum channels: A
characteristic function", Physical Review A \textbf{76}, 042331
(2007); arXiv:0707.4443.

\bibitem{Caruso} Caruso F., Eisert J., Giovannetti V., Holevo A.S., "Multi-mode bosonic Gaussian channels",
New Journal of Physics \textbf{10}, 083030 (2008); arXiv:0804.0511.

\bibitem{Caruso+} Caruso F., Eisert J., Giovannetti V., Holevo A.S., "The optimal unitary dilation for bosonic Gaussian channels",
Phys. Rev. A \textbf{84}, 022306 (2011); arXiv:1009.1108.

\bibitem{p-e-b-ch} Chruscinski D., Kossakowski A., "On Partially Entanglement Breaking Channels",
Open Sys. Information Dyn., \textbf{13}, 17-26 (2006);
arXiv:quant-ph/0511244.

\bibitem{DVV} Demoen B., Vanheuverzwijn P., Verbeure A., "Completely positive quasi-free
maps on the CCR algebra", Rep. Math. Phys. \textbf{15}, 27-39
(1979).

\bibitem{D&Sh} Devetak I., Shor P., "The capacity of a quantum channel for simultaneous transition of classical and quantum information",
arXiv:quant-ph/0311131.

\bibitem{W&Co} Duan R.,
Severini S., Winter A., "Zero-error communication via quantum
channels, non-commutative graphs and a quantum Lovasz theta
function", IEEE Trans. Inf. Theory \textbf{59}:2, 1164-1174 (2013).

\bibitem{E&W} Eisert J., Wolf M.M., "Gaussian quantum channels", Quantum Information with Continuous
Variables of Atoms and Light, 23-42 (Imperial College Press, London,
2007); arXiv:quant-ph/0505151.

\bibitem{H&Co} Hiai F., Mosonyi M., Petz D., Beny C., "Quantum f-divergences and error correction", Rev. Math. Phys. \textbf{23}:7,
691-747; arXiv:1008.2529.

\bibitem{H-BC} Holevo A.S., "Towards the mathematical theory of quantum communication
channels", Probl. Inform. Transm. \textbf{8}:1, 63-71 (1972).

\bibitem{H-SCI} Holevo A.S., \emph{Quantum systems, channels, information.
A mathematical introduction} (Berlin, DeGruyter, 2012).

\bibitem{H-EBC} Holevo A.S., "On extreme Bosonic linear channels",
arXiv:1111.3552.

\bibitem{H-1MGC} Holevo A.S., "One-mode quantum Gaussian channels:
structure and quantum capacity", Problems of Information
Transmission, \textbf{43}:1, 3-14 (2007); quant-ph/0607051.

\bibitem{H-c-c} Holevo A.S., "On complementary
channels and the additivity problem", Probability Theory and
Applications, \textbf{51}:1, 134-143 (2006); arXiv:quant-ph/0509101.

\bibitem{Ivan} Ivan J.S., Sabapathy K.K., Simon R., "Operator-sum Representation for Bosonic Gaussian Channels",
Phys. Rev. A \textbf{84}, 042311 (2011); arXiv:1012.4266.

\bibitem{J&P} Jencova A., Petz D., "Sufficiency in quantum statistical inference",
Commun. Math. Phys. \textbf{263}, 259-276 (2006);
arXiv:math-ph/0412093.

\bibitem{J-rev} Jencova A., "Reversibility conditions for quantum operations", Rev.Math.Phys. \textbf{24}, 1250016 (2012);
arXiv:1107.0453.

\bibitem{K&M} Kostrikin A.I., Manin Yu.I.,
\emph{Linear Algebra and Geometry} (CRC, 1989).

\bibitem{ZEC} Medeiros R.A.C., de Assis F.M. "Quantum zero-error capacity",
Int. J. Quant. Inf., \textbf{3}, 135 (2005).

\bibitem{N&Ch} Nielsen M.A., Chuang I.L., \emph{Quantum Computation and Quantum
Information} (Cambridge University Press, 2000).

\bibitem{O&Co} Ogawa T., Sasaki A., Iwamoto M., Yamamoto H.,
"Quantum Secret Sharing Schemes and Reversibility of Quantum
Operations", Phys. Rev. A \textbf{72}, 032318 (2005);
arXiv:quant-ph/0505001.

\bibitem{P-sqc} Petz D., "Sufficiency of channels over von Neumann algebras", Quart. J. Math. Oxford
Ser. (2) \textbf{39}:153, 97-108 (1988).

\bibitem{P&V} Plenio M.B., Virmani S., "An introduction to
entanglement measures",  Quantum Inf. Comput., \textbf{7}, 1-51
(2007); arXiv:quant-ph/0504163.

\bibitem{Sh-RC} Shirokov M.E., "Reversibility conditions for quantum channels and their applications",
Sbornic:Mathematics, \textbf{204}:8, 1215-1237 (2013);
arXiv:1203.0262.

\bibitem{Sh-SBC} Shirokov M.E., "On singular Bosonic linear channels", arXiv:1302.6879.


\bibitem{n-ch} A noiseless channel is unitary equivalent to the channel
$\rho\mapsto\rho\otimes\sigma$, where $\sigma$ is a given state; a
Bosonic linear channel $\,\Phi_{K,f}$ is noiseless if its parameters
satisfy the conditions: $\ker K$ is a symplectic subspace of $Z_B$
(possibly, trivial), $K|_{[\ker K]^{\perp}}$ is a symplectic
trasformation $[\ker K]^{\perp}\rightarrow Z_A$, $f(z)=\phi(z_1)$,
where $z_1$ is the first component of $z$ corresponding to the
decomposition $Z_B=\ker K\oplus[\ker K]^{\perp}$ and $\phi$ is a
characteristic function of a particular state.


\end{thebibliography}
\end{document}